\documentclass[12pt, preprint]{elsarticle}

\usepackage[usenames,dvipsnames]{xcolor}
\usepackage{lineno,hyperref}
\usepackage{graphics}
\usepackage{chemformula}
\usepackage[T1]{fontenc}
\usepackage{dcolumn}
\usepackage{subcaption}
\usepackage{cleveref}
\usepackage{times}
\usepackage{tablefootnote}
\usepackage{xcolor}
\usepackage{multirow}
\usepackage{siunitx}
\usepackage{mathtools}
\usepackage{bm}
\usepackage{esvect}
\usepackage{setspace}
\usepackage[margin=2.5cm]{geometry}
\usepackage{amssymb} 

\newcommand{\meth}{CH$_4$}
\newcommand{\hydro}{H$_2$}
\newcommand{\nitro}{N$_2$}

\newcommand{\degC}{$^{\circ}$C}
\newcommand{\cm}{cm$^{-1}$}

\newcommand{\RevCAR}[1]{{\color{blue}#1}}
\newcommand{\RevGLAS}[1]{{\color{ForestGreen}#1}}
\newcommand{\RevBRS}[1]{{\color{violet}#1}}
\newcommand{\RevDJW}[1]{{\color{brown}#1}}
\newcommand{\RevRAO}[1]{{\color{brown}#1}}

\newcommand{\RevCARJC}[1]{{\color{blue}#1}}
\newcommand{\CarbonA}[1]{{\color{NavyBlue}#1}}
\newcommand{\CarbonB}[1]{{\color{OliveGreen}#1}}
\newcommand{\CarbonC}[1]{{\color{WildStrawberry}#1}}

\renewcommand{\RevCAR}[1]{#1} 
\renewcommand{\RevGLAS}[1]{#1} 
\renewcommand{\RevBRS}[1]{#1} 
\renewcommand{\RevDJW}[1]{#1} 
\renewcommand{\RevRAO}[1]{#1} 
\renewcommand{\RevCARJC}[1]{#1} 

\renewcommand{\CarbonA}[1]{#1} 
\renewcommand{\CarbonB}[1]{#1} 
\renewcommand{\CarbonC}[1]{#1} 

\journal{Journal of \LaTeX\ Templates}

\makeatletter
\providecommand{\doi}[1]{%
  \begingroup
    \let\bibinfo\@secondoftwo
    \urlstyle{rm}%
    \href{http://dx.doi.org/#1}{%
      doi:\discretionary{}{}{}%
      \nolinkurl{#1}%
    }%
  \endgroup
}
\makeatother
\begin{document}
\begin{frontmatter}
\title{Thermal stress modelling of diamond on GaN/III-Nitride membranes}

\author[CU-PHYSX]{Jerome A. Cuenca\corref{mycorrespondingauthor}}\ead{cuencaj@cardiff.ac.uk}
\author[UOG-ENGIN]{Matthew D. Smith}
\author[UOB-PHYS]{Daniel E. Field}
\author[UOC-MAT,UOS-PHYS]{Fabien C-P. Massabuau}
\author[CU-PHYSX]{Soumen Mandal}
\author[UOB-PHYS]{James Pomeroy}
\author[UOC-MAT,CU-ENGIN]{David J. Wallis}
\author[UOC-MAT]{Rachel A. Oliver}
\author[UOG-ENGIN]{Iain Thayne}
\author[UOB-PHYS]{Martin Kuball}
\author[CU-PHYSX]{Oliver A. Williams}
\address[CU-PHYSX]{School of Physics and Astronomy, Cardiff University Cardiff, Wales, CF24 3AA, UK}
\address[UOG-ENGIN]{School of Engineering, University of Glasgow, G12 8LT, UK}
\address[UOB-PHYS]{Centre for Device Thermography and Reliability, University of Bristol, BS8 1TL, UK}
\address[UOC-MAT]{Cambridge Centre for Gallium Nitride, Department of Materials Science and Metallurgy, Cambridge, CB3 0FS, UK}
\address[UOS-PHYS]{Department of Physics, SUPA, University of Strathclyde, Glasgow, G4 0NG, UK}
\address[CU-ENGIN]{School of Engineering, Cardiff University Cardiff, Wales, CF24 3AA, UK}
\cortext[mycorrespondingauthor]{Corresponding author}

\setstretch{1.5}
\begin{abstract}
Diamond heat-spreaders for gallium nitride (GaN) devices currently depend upon a robust wafer bonding process. Bonding-free membrane methods demonstrate potential, however, chemical vapour deposition (CVD) of diamond directly onto a III-nitride (III-N) heterostructure membrane induces significant thermal stresses. In this work, these thermal stresses are investigated using \CarbonC{an} analytical \CarbonC{approach}, a numerical \CarbonC{model} and experimental validation. The thermal stresses are caused by the mismatch in the coefficient of thermal expansion (CTE) between the GaN/III-N stack, silicon (Si) and the diamond from room temperature to CVD growth temperatures. Simplified analytical wafer bow models underestimate the membrane bow \CarbonC{for small sizes} while numerical models replicate the stresses and bows with increased accuracy using temperature gradients. The largest tensile stress measured using Raman spectroscopy at room temperature was approximately 1.0 $\pm0.2$ GPa while surface profilometry shows membrane bows as large as \SI{58}{\micro\metre}. This large bow is caused by additional stresses from the Si frame in the initial heating phase \CarbonC{which are held in place by the diamond} and highlights challenges for any device fabrication using contact lithography. However, the bow can be reduced if the membrane is pre-stressed to become flat at CVD temperatures. In this way, a sufficient platform to grow diamond on GaN/III-N structures without wafer bonding can be realised.
\end{abstract}

\begin{keyword}
cvd diamond, gallium nitride, membranes, thermal stress, finite element modelling
\end{keyword}
\end{frontmatter}


\setstretch{1.5}
\section{Introduction}
Increasing demand for high power microwave devices has driven research into gallium nitride (GaN), in particular, high electron mobility transistors (HEMTs)\cite{Ejeckam2014}. Current device technologies either have impressive high power handling (e.g. Si) or high frequency capability (e.g. GaAs and InP), though not both. GaN is a possible candidate although, despite significant progress, its power handling capability is still limited by self-heating\cite{Ejeckam2014}. This is due to the limited thermal conductivity of GaN with values between 160 and 210 W/mK\cite{Zheng2019}. Additionally, some substrates for GaN also have a low thermal conductivity such as sapphire (Al$_2$O$_3$) and silicon (Si) at around 30 and 150 W/mK, respectively\cite{Zheng2019,Pishchik2009,Pomeroy2014}. This can be overcome with sufficient heat extraction by integration with substrates of much higher thermal conductivity such as silicon carbide (SiC) at 400 W/mK, which is commonly used for electronic applications. Diamond is another candidate which in its single crystal form is capable of achieving thermal conductivities in excess of 2,000 W/mK\cite{Wei1993, Hirama2011}. Several studies have shown reduced operating temperatures of GaN devices when using integrated diamond heat spreading technology\cite{Tadjer2012,Zhou2017,Han2015}.

GaN-on-diamond is not trivial to fabricate, but can be realised either by surface activated wafer bonding\cite{Gerrer2018,Francis2010,Mu2018,Wang2020,Kim2018} or heteroepitaxial growth. Growth of GaN onto diamond\cite{Zhang2010} or diamond onto GaN can be achieved using interlayers such as silicon nitride (SiN) or group-III-Nitride (III-N)  structures\CarbonC{. Examples of III-N structures include} aluminium nitride (AlN) and aluminium gallium nitride (AlGaN)\cite{Sun2015b,Smith2020a,Hirama2011,Hetzl2018,Mandal2017}. In the wafer bonding approach, it is imperative to achieve a sub-nanometre surface roughness in order for Van der Waals bonding to occur. However, it has been calculated that a large thermal barrier resistance of Van der Waals bonded diamond on GaN is expected, reducing the efficacy of diamond heat spreading technology\cite{Waller2020}. For heteroepitaxial growth of diamond onto III-N structures, this can only be achieved at present with polycrystalline diamond (PCD) using chemical vapour deposition (CVD) at lower than atmospheric pressure and at high temperatures (0.1 to 0.2 atm at temperatures from 600 to 800 \degC{}\cite{Zhou2017}). Additionally in contrast to single crystal diamond (SCD), PCD can be grown over  large areas, making this approach more commercially viable. A number of CVD methods exist for depositing PCD. Microwave plasma assisted chemical vapour deposition (MPCVD) offers diamond growth over small areas (<100 mm diameter wafers) and can be optimised for very fast growth rates (50 to \SI{150}{\micro\metre} h$^{-1}$)\cite{Yan2002}. Larger area coverage (>100 mm diameter wafers) can be achieved with hot filament (HFCVD), although repeatability is challenging and growth rates are generally much lower (\CarbonA{1 to \SI{4}{\micro\metre} h$^{-1}$}) \cite{Ali2011, Barber1997, Amaral2006}.

\begin{figure}[t!]
	\centering
	\resizebox{12cm}{!}{
	\begin{tikzpicture}
 
\node [shape=rectangle, minimum width=1cm, minimum height=0.5cm, text height=0.25cm, anchor=west] 	at (-1.5,1.5) {\Large\bf Wafer Bonding Method};
\node [shape=rectangle, minimum width=1cm, minimum height=0.5cm, text height=0.25cm, anchor=west] 	at (-1.5,-5.5) {\Large\bf Membrane Method};
   
\node[anchor=center] at (0,0) {\includegraphics[width=3cm]{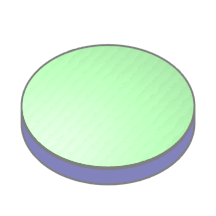}};
\node[anchor=center] at (4,0) {\includegraphics[width=3cm]{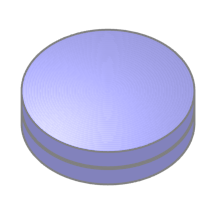}};
\node[anchor=center] at (8,0) {\includegraphics[width=3cm]{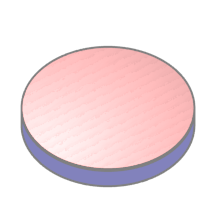}};
\node[anchor=center] at (12,0) {\includegraphics[width=3cm]{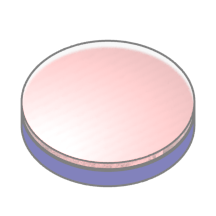}};
\node[anchor=center] at (16,0) {\includegraphics[width=3cm]{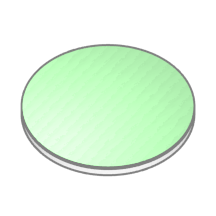}};

\node [draw=gray, thick, fill=green!20!white, shape=rectangle, minimum width=3cm, minimum height=0.5cm, text height=0.25cm, anchor=center] 	at (0,-3) {GaN};
\node [draw=gray, thick, fill=red!20!white, shape=rectangle, minimum width=3cm, minimum height=0.5cm, text height=0.25cm, anchor=center] 		at (0,-3.5) {AlN/AlGaN};
\node [draw=gray, thick, fill=blue!20!white, shape=rectangle, minimum width=3cm, minimum height=0.5cm, text height=0.25cm, anchor=center] 	at (0,-4) {Si};

\node [draw=gray, thick, fill=blue!20!white, shape=rectangle, minimum width=3cm, minimum height=0.5cm, text height=0.25cm, anchor=center] 	at (4,-2) {Si Temp};
\node [draw=gray, thick, fill=orange!20!white, shape=rectangle, minimum width=3cm, minimum height=0.5cm, text height=0.25cm, anchor=center] 	at (4,-2.5) {Oxide/Nitride};
\node [draw=gray, thick, fill=green!20!white, shape=rectangle, minimum width=3cm, minimum height=0.5cm, text height=0.25cm, anchor=center] 	at (4,-3) {GaN};
\node [draw=gray, thick, fill=red!20!white, shape=rectangle, minimum width=3cm, minimum height=0.5cm, text height=0.25cm, anchor=center] 		at (4,-3.5) {AlN/AlGaN};
\node [draw=gray, thick, fill=blue!20!white, shape=rectangle, minimum width=3cm, minimum height=0.5cm, text height=0.25cm, anchor=center] 	at (4,-4) {Si};

\node [draw=gray, thick, fill=red!20!white, shape=rectangle, minimum width=3cm, minimum height=0.5cm, text height=0.25cm, anchor=center] 		at (8,-2.5) {AlN/AlGaN};
\node [draw=gray, thick, fill=green!20!white, shape=rectangle, minimum width=3cm, minimum height=0.5cm, text height=0.25cm, anchor=center] 	at (8,-3) {GaN};
\node [draw=gray, thick, fill=orange!20!white, shape=rectangle, minimum width=3cm, minimum height=0.5cm, text height=0.25cm, anchor=center] 	at (8,-3.5) {Oxide/Nitride};
\node [draw=gray, thick, fill=blue!20!white, shape=rectangle, minimum width=3cm, minimum height=0.5cm, text height=0.25cm, anchor=center] 	at (8,-4) {Si Temp};

\node [draw=gray, thick, fill=gray!20!white, shape=rectangle, minimum width=3cm, minimum height=0.5cm, text height=0.25cm, anchor=center] 	at (12,-2) {Diamond};
\node [draw=gray, thick, fill=red!20!white, shape=rectangle, minimum width=3cm, minimum height=0.5cm, text height=0.25cm, anchor=center] 		at (12,-2.5) {AlN/AlGaN};
\node [draw=gray, thick, fill=green!20!white, shape=rectangle, minimum width=3cm, minimum height=0.5cm, text height=0.25cm, anchor=center] 	at (12,-3) {GaN};
\node [draw=gray, thick, fill=orange!20!white, shape=rectangle, minimum width=3cm, minimum height=0.5cm, text height=0.25cm, anchor=center] 	at (12,-3.5) {Oxide/Nitride};
\node [draw=gray, thick, fill=blue!20!white, shape=rectangle, minimum width=3cm, minimum height=0.5cm, text height=0.25cm, anchor=center] 	at (12,-4) {Si Temp};

\node [draw=gray, thick, fill=gray!20!white, shape=rectangle, minimum width=3cm, minimum height=0.5cm, text height=0.25cm, anchor=center] 	at (16,-4) {Diamond};
\node [draw=gray, thick, fill=red!20!white, shape=rectangle, minimum width=3cm, minimum height=0.5cm, text height=0.25cm, anchor=center] 		at (16,-3.5) {AlN/AlGaN};
\node [draw=gray, thick, fill=green!20!white, shape=rectangle, minimum width=3cm, minimum height=0.5cm, text height=0.25cm, anchor=center] 	at (16,-3) {GaN};

\draw[->,line width=1mm] (1.5,0) -- (2.5,0);
\draw[->,line width=1mm] (5.5,0) -- (6.5,0);
\draw[->,line width=1mm] (9.5,0) -- (10.5,0);
\draw[->,line width=1mm] (13.5,0) -- (14.5,0);

\node [shape=rectangle, minimum width=1cm, minimum height=0.5cm, text height=0.25cm, anchor=west] 	at (5.5,0.5) {\Large$\circlearrowleft$};
\node [shape=rectangle, minimum width=1cm, minimum height=0.5cm, text height=0.25cm, anchor=west] 	at (13.5,0.5) {\Large$\circlearrowleft$};

\node[anchor=center] at (0,-6.75) {\includegraphics[width=4cm]{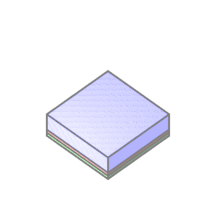}};
\node[anchor=center] at (4,-6.75) {\includegraphics[width=4cm]{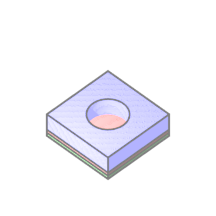}};
\node[anchor=center] at (8,-6.75) {\includegraphics[width=4cm]{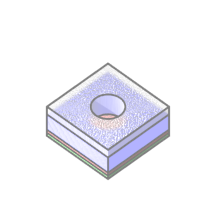}};

\node [draw=gray, thick, fill=red!20!white, shape=rectangle, minimum width=3cm, minimum height=0.5cm, text height=0.25cm, anchor=center] 		at (0,-10) {AlN/AlGaN};
\node [draw=gray, thick, fill=blue!20!white, shape=rectangle, minimum width=3cm, minimum height=0.5cm, text height=0.25cm, anchor=center] 	at (0,-9.5) {Si};
\node [draw=gray, thick, fill=green!20!white, shape=rectangle, minimum width=3cm, minimum height=0.5cm, text height=0.25cm, anchor=center] 	at (0,-10.5) {GaN};

\node [draw=gray, thick, fill=green!20!white, shape=rectangle, minimum width=3cm, minimum height=0.5cm, text height=0.25cm, anchor=center]	at (4,-10.5) {GaN};
\node [draw=gray, thick, fill=blue!20!white, shape=rectangle, minimum width=1cm, minimum height=0.5cm, text height=0.25cm, anchor=center] 	at (5,-9.5) {Si};
\node [draw=gray, thick, fill=blue!20!white, shape=rectangle, minimum width=1cm, minimum height=0.5cm, text height=0.25cm, anchor=center] 	at (3,-9.5) {Si};
\node [draw=gray, thick, fill=red!20!white, shape=rectangle, minimum width=3cm, minimum height=0.5cm, text height=0.25cm, anchor=center] 		at (4,-10) {AlN AlGaN};

\node [draw=gray, thick, fill=green!20!white, shape=rectangle, minimum width=3cm, minimum height=0.5cm, text height=0.25cm, anchor=center]	at (8,-10.5) {GaN};
\node [draw=gray, thick, fill=blue!20!white, shape=rectangle, minimum width=1cm, minimum height=0.5cm, text height=0.25cm, anchor=center] 	at (9,-9.5) {Si};
\node [draw=gray, thick, fill=blue!20!white, shape=rectangle, minimum width=1cm, minimum height=0.5cm, text height=0.25cm, anchor=center] 	at (7,-9.5) {Si};
\node [shape=rectangle, minimum width=1cm, minimum height=0.5cm, text height=0.25cm, anchor=west] 	at (8,-8.5) {Diamond};
\draw[draw=gray, thick, fill=gray!20!white] (6.5,-8.75) -- (7.75,-8.75) -- (7.75,-9.25) -- (8.25,-9.25) -- (8.25,-8.75) -- (9.5,-8.75) -- (9.5,-9.25) -- (8.5,-9.25) -- (8.5,-10.25) -- 
	(7.5,-10.25) -- (7.5,-9.25) -- (6.5,-9.25) 
	-- cycle;
\node [draw=gray, thick, fill=red!20!white, shape=rectangle, minimum width=3cm, minimum height=0.5cm, text height=0.25cm, anchor=center] 		at (8,-10) {AlN AlGaN};

\draw[->,line width=1mm] (1.5,-7) -- (2.5,-7);
\draw[->,line width=1mm] (5.5,-7) -- (6.5,-7);

\end{tikzpicture}
	}
	\caption{Approaches to diamond on GaN/III-N integration using wafer bonding steps prior to CVD diamond growth (top) and membrane fabrication (bottom). `Si Temp' is defined as a temporary Si carrier wafer.}
	\label{fig:bond}
\end{figure}
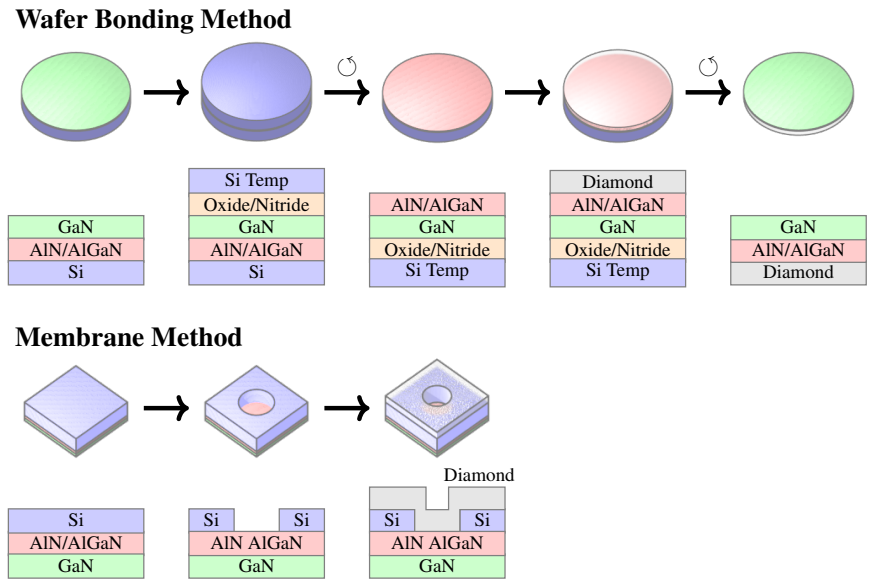
Direct diamond growth on GaN is challenging owing to a number of factors. The first is that a stable carbide bond is difficult to achieve on either side of the GaN faces which can be  overcome using interlayers that form carbides. The second issue is that GaN is widely deposited using metal organic chemical vapour deposition (MOCVD) where the Ga-polar HEMT layer is typically at the top of the stack. While growing diamond on the Ga-polar side has been demonstrated for thin films at less than \SI{1}{\micro\metre}\cite{Tadjer2012}, this creates additional challenges to embed the diamond layer during device fabrication. Also, it has been shown that the thermal conductivity of diamond can be very low at these thicknesses\cite{Zhou2017}. Therefore, the diamond must ideally be grown on the underside or the N-polar side of the GaN/III-N stack. The N-polar side is chemically very different to the Ga-polar surface and thick film growth requires significant investigation\cite{Mandal2017}. This requires wafer bonding to a temporary carrier, flipping, etching and then \CarbonC{growing the diamond on the exposed III-N stack} as is shown in Fig. \ref{fig:bond}. Finally, a prominent issue is the thermal stress and strain (bow) from the mismatch in the coefficient of thermal expansion (CTE) between the layers in the stack\cite{Edwards2010}. High temperature CVD diamond growth generates large stresses upon cooling which bows the wafer and may mechanically fracture the GaN. Additionally, these high temperatures may also result in complete destruction of the GaN/III-N stack and so ensuring the growth temperature is as low as possible mitigates the risk of damage or fracturing upon cooling.

A potential method which enables fast access to the correct polarity face of the GaN/III-N stack for diamond growth, without the need for costly and time consuming wafer bonding, is to use a GaN/III-N membrane, as is shown in Fig. \ref{fig:bond}. This can be achieved using an inductively coupled plasma (ICP)\cite{Smith2020a}. This membrane technique has been realised for other GaN applications including light emitting diodes, planar photonic devices\cite{Li2015b,Wang2014,Liu2018} and impedance based sensors\cite{Kang2005a, Lalinsky2012, Alifragis2016}. The details of the integration of diamond with GaN membranes are seldom reported, specifically owing to the fragile nature of the membrane and the high temperatures required to grow diamond. Previous reports have shown diamond on GaN/III-N membranes from 0.5 to 5 mm in diameter under stresses of around 0.7 GPa due to the thermal cycling \cite{Smith2020a}. The stress is lower than the experimentally reported mechanical strength of GaN (compressive strength of 10 to 15 GPa and tensile strength of 4 to 7.5 GPa)\cite{Sung2010,Nowak1999,Brown2011} and far less than density functional theory (DFT) estimates of the tensile strength (30 to 40 GPa)\cite{Ahn2015,Umeno2015}. The stresses are also equivalent to those measured in a typical diamond on GaN/III-N sample produced through the wafer bonding method (0.67 to 1 GPa)\cite{Hancock2016}, demonstrating a potentially viable \CarbonC{approach} for circumventing wafer bonding. However, the membrane deformations and bow are not reported which may significantly impact device fabrication using contact lithography. Since the membrane structure is far more complex when compared to a cylindrical wafer stack, an understanding of the parameters which cause these deformations and induced thermal stresses is required before considering this approach for any device fabrication.

In this work, the thermal stress and strains (membrane bow) associated with the growth of CVD diamond on GaN/III-N membranes have been investigated using analytical modelling, numerical modelling and experimental work. First, \CarbonC{a simplified} Stoney analytical model is briefly described for stress in a two layer cylindrical stack. Following this a finite element model (FEM) is presented which calculates the anticipated thermal stresses in circular membranes and investigates the effect of varying the geometry, materials and temperature. Next, experimental data is presented on the growth and characterisation of diamond on GaN/III-N membranes to validate the model. Finally, comparisons and discussions are drawn, particularly on the limitations of the CVD diamond on GaN/III-N membrane approach.

\section{Analytical Model\label{sec:theory}}

\subsection{Coefficient of Thermal Expansion (CTE)}

\RevCARJC{Thin film deposition at elevated temperatures requires careful consideration of stress. At room temperature, the net stresses are due to a combination of intrinsic stresses and extrinsic stresses. Intrinsic stresses arise mainly due to the imperfections in the microstructure of the deposited film such as crystal grain boundaries, voids, impurities and defects caused by the lattice mismatch between the film and the substrate. For example, the lattice mismatch between GaN and Si results in residual intrinsic stresses after MOCVD\cite{Pan2011}. Additionally, it is well-known that intrinsic stresses are also present in CVD diamond depending on the experimental conditions\cite{Windischmann1991,Rajamani2004}. Extrinsic stresses are due to externally applied forces such as mechanical loading or thermal stress, the latter of which is a well-known challenge for thin film deposition and is the subject of this study.}

\begin{figure}[h!]
	\centering
	\includegraphics[width=0.5625\textwidth]{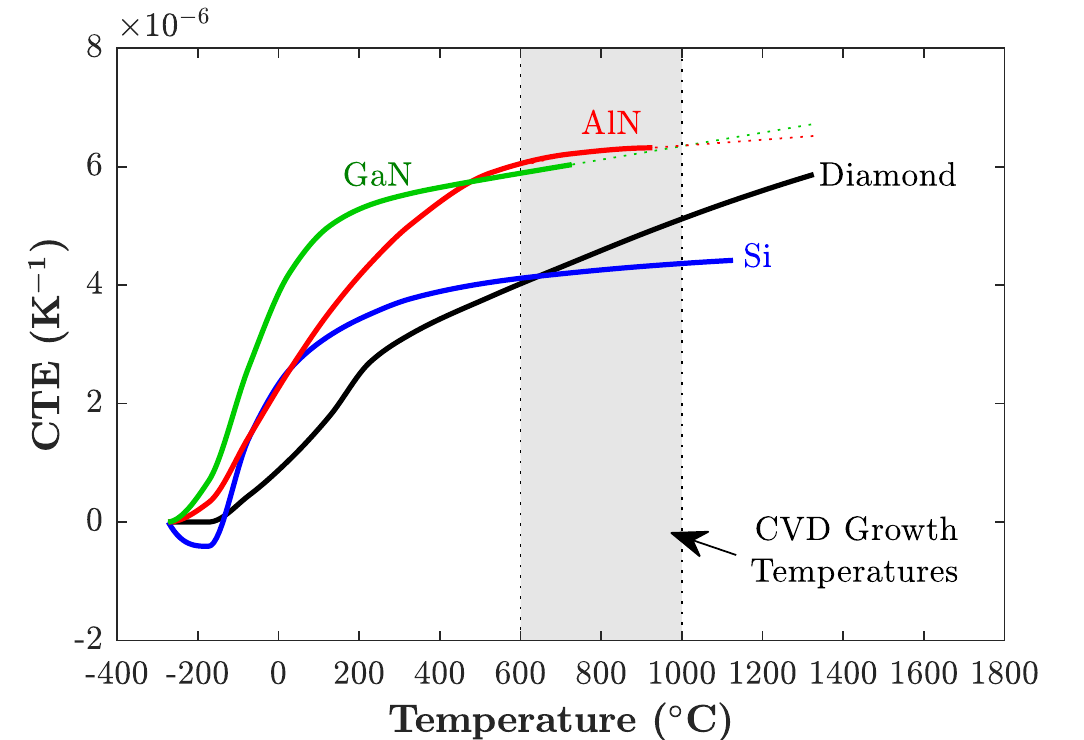}
	\caption{Temperature dependent CTE of diamond, Si, AlN and GaN\cite{Edwards2010,Slack1975,Roder2005}. }
	\label{fig:cte}
\end{figure}

Thermal stresses arise due to the thermomechanical properties of the layers in the stack. In isolation, each material preferentially expands/contracts to different volumes at different temperatures. However, for a three dimensional (3D) sample with all layers bound axially in the $z$ direction, the deformations in the $xy$ plane are constrained. The CTE of \CarbonC{diamond and the layers in a typical GaN-on-Si wafer} are  shown in Fig.  \ref{fig:cte}. AlN and GaN have very similar values at typical CVD growth temperatures of 600 to 1000 \degC{} and a small difference at room temperature of around 1.5 $\times 10^{-6}$ \si{\kelvin^{-1}}. Diamond and Si have lower values than the GaN/III-N layers across the temperature range. At room temperature Si intersects with AlN while diamond is much lower than all of the materials. It is therefore expected that after cooling to room temperature, the CVD diamond on the membrane will be held in compression, whilst the GaN and III-N layers in tension. Thus, careful consideration of the \textit{tensile} strength of the GaN/III-N stack must be made to ensure that the membrane does not fracture through upon cooling. \RevCARJC{The resultant deformations caused by thermal stress can be determined for a simplified cylindrical wafer stack using the following analytical model.}

\subsection{Analytical Model}
In a cylindrical coordinate system ($\rho,\theta,z$), consider an axially symmetric wafer stack consisting of two unbound layers of differing thicknesses ($t_{\textrm{film}}$ and $t_{\textrm{sub}}$) with differing CTE ($\alpha_{\textrm{film}}$ and $\alpha_{\textrm{sub}}$). The layer thicknesses in the $z$ direction are much smaller than the radius of the cylindrical stack. When the temperature is lowered, depending on the CTE, the volumes of the layers shrink with a negligible change in thickness compared to a much larger radial change. Using the linear thermal expansion model, the resultant radial 1D strains are:

\begin{equation}
	  \varepsilon_{\textrm{film}} = \frac{r_{\textrm{film}}(T)}{r_{\textrm{0,film}}} = 1+\alpha_{\textrm{film}}(T) \left[T-T_0\right]
	  \label{eq:strainf}
\end{equation}
\begin{equation}
	  \varepsilon_{\textrm{sub}} = \frac{r_{\textrm{sub}}(T)}{r_{\textrm{0,sub}}} = 1+\alpha_{\textrm{sub}}(T) \left[T-T_0\right]
	\label{eq:strains}
\end{equation}

\noindent where $T$ is the temperature in K, $r(T)$ is the temperature dependent radius of the layer in m, $r_0$ denotes the initial radius in m at a reference temperature of $T_0$ and $\alpha(T)$ is the temperature dependent CTE of the layer in K$^{-1}$. \RevGLAS{If the layers are bound, the radial strain at the interface must be equal, with one layer being held in tension while the other is held in compression}. Therefore, an additional strain is required to keep the materials bound at the interface, equal to the difference between the two unbound strains:

\begin{equation}
	\Delta\varepsilon = \varepsilon_{\textrm{sub}} - \varepsilon_{\textrm{film}}= \left[T-T_0\right]\left[ \alpha_{\textrm{sub}}(T) - \alpha_{\textrm{film}}(T)\right] = \Delta T \Delta\alpha
\end{equation}

\noindent {The calculated stress associated with this strain is related through the Young's modulus:}

\begin{equation}
	\sigma  = E\cdot \Delta T \Delta \alpha
	\label{eq:thermalstress}
\end{equation}

\noindent where $\sigma$ is the induced thermal stress in Pa and $E$ is the Young's modulus of the layer in Pa. Considering 2D axial symmetry however, this generated 1D radial stress is not the equilibrium state as the two layer system is not restricted along the $z$ axis. The induced internal stress thus results in a finite strain in the $z$ direction. For an axially symmetric sample, this results in a radius of curvature $R$ through a relation with the biaxial modulus, otherwise known as the Stoney formula\cite{Windischmann1991,Seshan2001,Franssila2010}:

\begin{equation}
 	R = \frac{E_{\textrm{sub}}}{\left(1-\nu_{\textrm{sub}}\right)}   \frac{t_{\textrm{sub}}^2}{t_{\textrm{film}}} \frac{1}{6} \frac{1}{\sigma}
 	\label{eq:stoney}
\end{equation}

\noindent where $R$ is the radius of curvature of the substrate, $E_{\textrm{sub}}/\left(1-\nu_{\textrm{sub}}\right)$ is the biaxial modulus of the substrate and \RevGLAS{$\nu_{\textrm{sub}}$} is the Poisson's ratio. With the radius of  curvature known, the wafer bow can be calculated for a given wafer diameter. This is simply related through arclength\cite{Edwards2010}:

\begin{equation}
	w = R \left( 1-\textrm{cos}\left[\frac{r_{\textrm{sub}}}{R}\right] \right)
\end{equation}

\noindent where $w$ is the displacement of the central point in the $z$ direction of the wafer in m or bow. Knowing the diameter of the wafer ( $2r_{\textrm{sub}}$), one can predict the radius of curvature of a cylindrical stack for a given thermal stress.

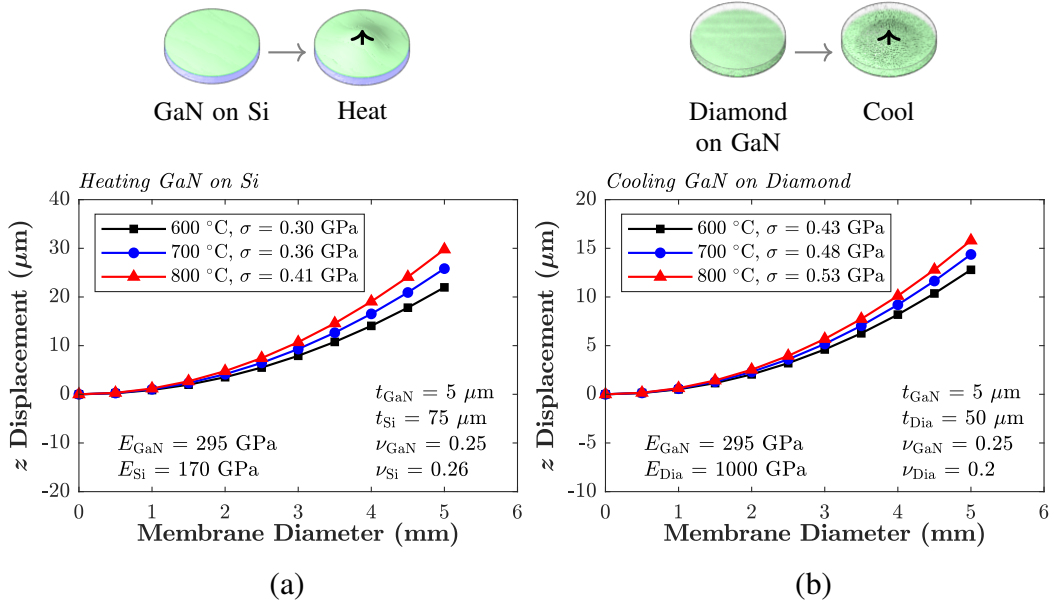
\begin{figure}[t!]\footnotesize
 	\setlength\tabcolsep{1pt}\centering
 	
	\begin{tikzpicture}\footnotesize
		\node[anchor=center,inner sep=0] at (0,-4.2) {\includegraphics[width=0.45\textwidth]{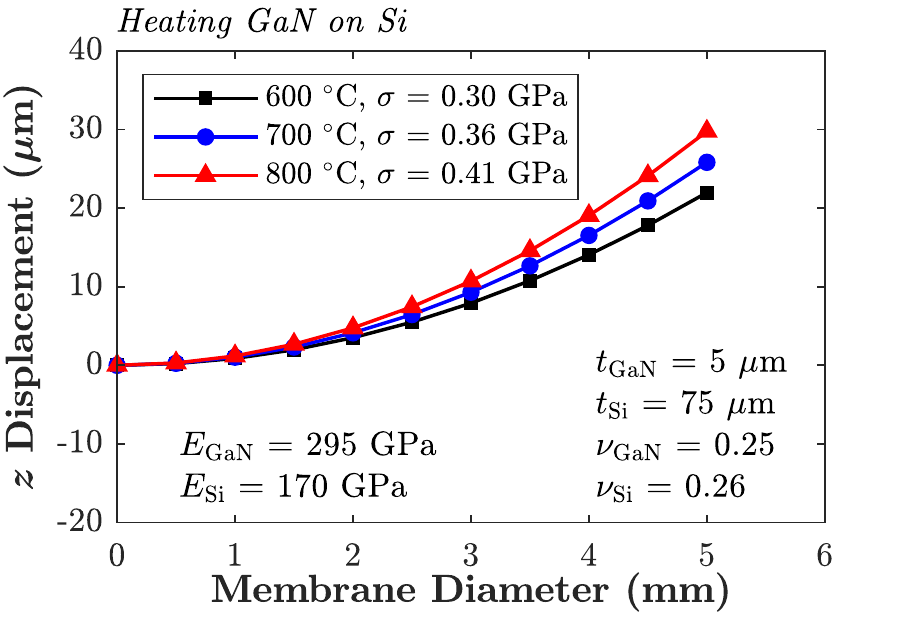}};
		\node[anchor=center,inner sep=0] at (7,-4.2) {\includegraphics[width=0.45\textwidth]{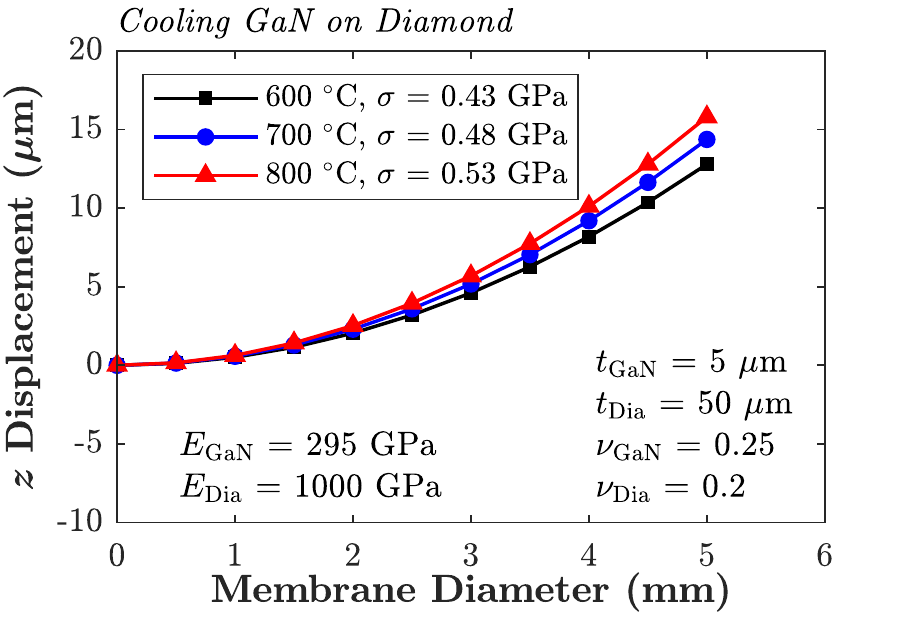}};
		\node[anchor=center,inner sep=0] at (-1,-0.1) {\includegraphics[trim=0 0 0 0,width=1.875cm]{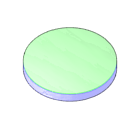}};
		\node[anchor=center,inner sep=0] at (1,-0.1) {\includegraphics[trim=0 0 0 0,width=1.875cm]{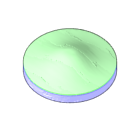}};
		\node[anchor=center,inner sep=0] at (6,-0.1) {\includegraphics[trim=0 0 0 0,width=1.875cm]{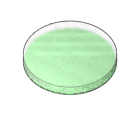}};
		\node[anchor=center,inner sep=0] at (8,-0.1) {\includegraphics[trim=0 0 0 0,width=1.875cm]{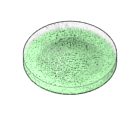}};
		\node[anchor=north,inner sep=0, text width=2cm, align=center] at (-1,-0.75) {GaN on Si};
		\node[anchor=north,inner sep=0, text width=2cm, align=center] at (1,-0.75) {Heat};
		\node[anchor=north,inner sep=0, text width=2cm, align=center] at (6,-0.75) {Diamond on GaN};
		\node[anchor=north,inner sep=0, text width=2cm, align=center] at (8,-0.75) {Cool};
		
		\draw[->, anchor=east, thick, gray] (-0.25,-0.1) -- ++ (0.5,0);
		\draw[->, anchor=east, thick, gray] (6.75,-0.1) -- ++ (0.5,0);	
		\draw[->, anchor=center, very thick, black] (1,0) -- ++ (0,0.2);
		\draw[->, anchor=center, very thick, black] (8,0) -- ++ (0,0.2);
		\normalsize
		\node[anchor=center,inner sep=0, text width=10cm, align=center] at (0,-7.2) {(a) };
		\node[anchor=center,inner sep=0, text width=10cm, align=center] at (7,-7.2) {(b)};
	\end{tikzpicture}
	\caption{Analytical model of  bow using Eq. (\ref{eq:stoney}) for two scenarios: \CarbonC{(a) the heating of a cylindrical GaN on Si stack to CVD growth temperatures, resulting in an upward bow} and (b) the cooling of a cylindrical diamond on GaN stack from CVD growth temperatures to room temperature, also resulting in an upward bow.}
  \label{fig:ana}
\end{figure}

The stresses in a cylindrical stack that are associated with the CVD diamond growth process at various deposition temperatures can be approximated using typical thicknesses, diameters and values of $E$, $\nu$ and $\alpha(T)$ for GaN and diamond as given in Fig. \ref{fig:cte} and Table \ref{tab:fem}. Fig. \ref{fig:ana} shows models of GaN-on-Si wafers heated to CVD growth temperatures and diamond on GaN wafers \RevRAO{cooled to room temperature}. Here, the GaN/AlGaN and AlN \RevDJW{nucleation} layers are \RevRAO{grouped and modelled as one thin GaN layer} ($t_{\textrm{film}}$= \SI{5}{\micro\meter}) while the substrate is modelled as Si  upon heating ($t_{\textrm{sub}}$= \SI{75}{\micro\meter}) or diamond on cooling ($t_{\textrm{sub}}$= \SI{50}{\micro\meter}). \RevDJW{With a constant radius of curvature, }it is clear that the bow increases with diameter\CarbonC{. At temperatures between 600 and 800 \degC{}, the initial heating bow is associated with a stress of around 0.3 to 0.41 GPa while the cooling bow is associated with a stress around of 0.43 to 0.53 GPa. An important note is that the analytical model implies that membranes of 0.5 mm in diameter have a bow of less than \SI{1}{\micro\metre}}. Low temperature growths (<800 \degC{}) are chosen in this study since most sucessful growths on GaN occur at lower temepratures to minimise thermal stress. This analytical model predicts a net stress in the GaN as high as \CarbonC{0.94 GPa}. \RevRAO{This value is lower than the reported tensile strength of GaN (4 to 7.5 GPa at room temperature) and similar to those reported in previous studies}\cite{Smith2020a, Brown2011}.

\begin{table}[t!]
	\centering
	\caption{Material properties used in analytical and numerical models.}
	\label{tab:fem}
	\footnotesize\noindent
	\begin{tabular}{lllllll}
	         \bf Material	& \bf	Young's Modulus			&	\bf Poisson's		&	\bf Density						&	\bf Thermal						&	\bf Specific 	&	\bf Ref.\\
	         {}			& \boldmath{$E$} \bf {(GPa)}		&	\bf Ratio			&	\bf \boldmath$\rho$ (g/m$^3$)	&	\bf Conductivity					&	\bf Heat		&	{}	\\	
	         {}			& {}								&	\boldmath $\nu$	&	{}								&	\bf\boldmath{$\kappa$} (W/mK)	&	\bf\boldmath{$C_p$ (J/kg$\cdot$K)}&	{}	\\
	         \hline
		Diamond 	& 1000							&	0.2		&	3.51		&	2270		&	0.52		&	\cite{Ralchenko2011,Mohr2014}\\
		GaN		& 295							&	0.25		&	6.16		&	253       		&	0.42		&	\cite{Nowak1999}		\\
		AlN			& $310.2-0.02T\textrm{e}^{-533T}$	&	0.23		&	2.92		&	321        		&	0.78		&	\cite{Bruls2001}	\\
		Si          		& $172.3-0.01T$	        			&	0.26		&	2.33		&	130			&	0.70        	&	\cite{Edwards2010,Slack1975}   	\\
	\end{tabular}
\end{table}

 A prominent limitation \RevBRS{of this model is its simplicity;} the membrane structure in reality is not a millimetre sized GaN-on-Si wafer but a millimetre sized GaN membrane that is suspended by a much larger Si border. The stresses induced by the Si cannot be taken into account since it is bound to both the edges of the membrane and the rest of the GaN film. \CarbonA{It is also assumed that the diamond thickness is constant on the top of the membrane}, which is not true since lateral growth will occur on the edges of the Si. Such model complexities can be readily implemented using a numerical model.

\section{Numerical Model}\label{sec:mod}
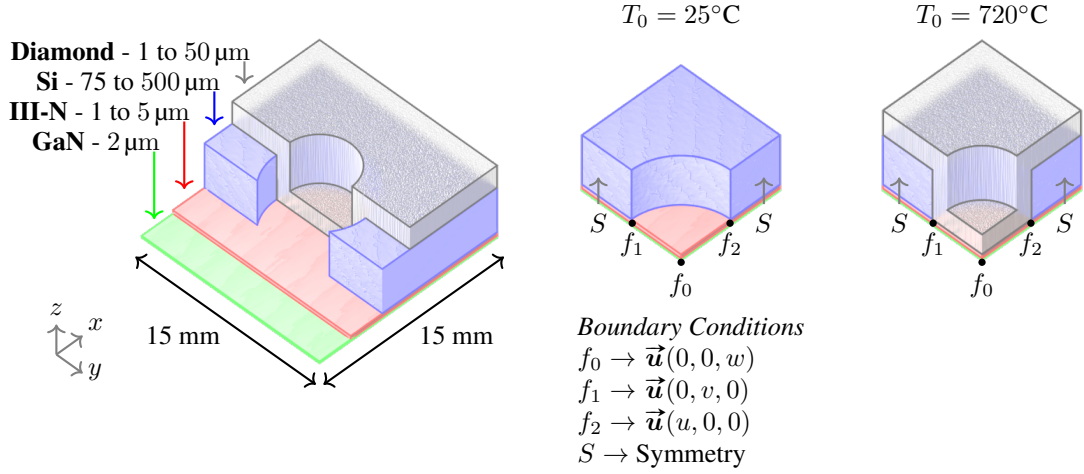
\begin{figure}[h!]
	\centering
	\begin{tikzpicture}[scale=1]\footnotesize
	\node[anchor=center,inner sep=0] at (0,0) {\includegraphics[width=5cm, trim=0 0 0 1cm]{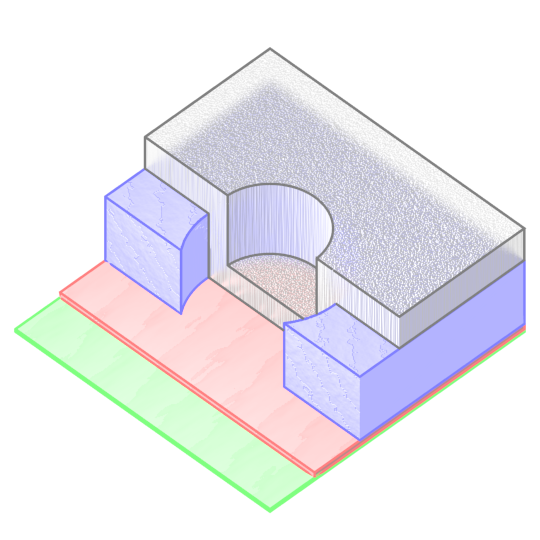}};
	\node[anchor=east] at (-0.8,2) (dia) {\textbf{Diamond} - 1 to \SI{50}{\micro\meter}};
	\node[anchor=east] at (-1.2,1.6) (si) {\textbf{Si} - 75 to \SI{500}{\micro\meter}};
	\node[anchor=east] at (-1.6,1.2) (aln) {\textbf{III-N} - 1 to \SI{5}{\micro\meter}};
	\node[anchor=east] at (-2,0.8) (gan) {\textbf{GaN} - \SI{2}{\micro\meter}};
	\draw[->, anchor=east, thick, gray] (-1.0,1.9) -- ++ (0,-0.3);
	\draw[->, anchor=east, thick, blue] (-1.4,1.5) -- ++ (0,-0.45);
	\draw[->, anchor=center, thick, red] (-1.8,1.1) -- ++ (0,-0.9);
	\draw[->, anchor=east, thick, green] (-2.2,0.7) -- ++ (0,-0.9);
	
	\draw[<->, anchor=east, thick, black] (-2.45,-0.7) -- ++ (2.4,-1.7);
	\draw[<->, anchor=east, thick, black] (2.45,-0.7) -- ++ (-2.4,-1.7);
	\node[anchor=north east] at (-1.2,-1.5) (wida) {15 mm};
	\node[anchor=north west] at (1.2,-1.5) (widb) {15 mm};
	
	\node[anchor=south] at (-3.5,-1.6) {$z$};
	\node[anchor=south west] at (-3.2,-1.8) {$x$};
	\node[anchor=north west] at (-3.2,-2) {$y$};
	\draw[->, anchor=center, thick, gray] (-3.5,-2) -- ++ (0,1.6/4);
	\draw[->, anchor=center, thick, gray] (-3.5,-2) -- ++ (1.4/4,1/4);
	\draw[->, anchor=center, thick, gray] (-3.5,-2) -- ++ (1.4/4,-1/4);
	\node[anchor=center] at (0,-3.5) { };
\end{tikzpicture}
\begin{tikzpicture}[scale=1]\footnotesize
	\node[anchor=center,inner sep=0] (g2) at (0,0) {\includegraphics[width = 4.4cm, trim=2cm 2.8cm 2cm 0cm]{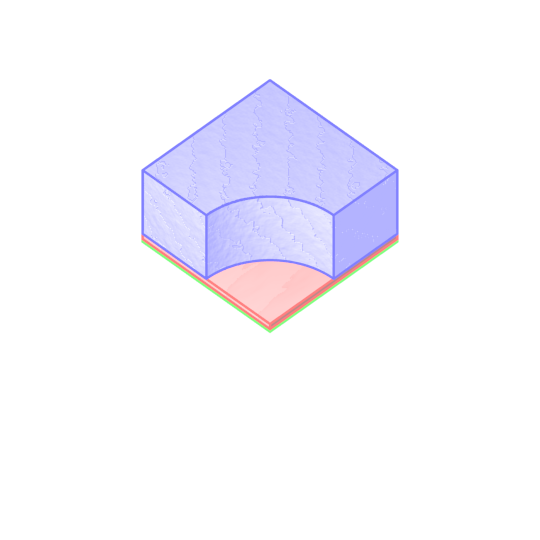}};

 	\path (g2) ++ (+1.1,-0.3) node (S1) [anchor=north ] {$S$};
 	\path (g2) ++ (-1.1,-0.3) node (S2) [anchor=north ] {$S$};
	\draw[->, anchor=center, thick, gray] (S1) -- ++ (0,0.6);
	\draw[->, anchor=center, thick, gray] (S2) -- ++ (0,0.6);
    
 	\path (g2) ++ (+0.65,-0.38) node (nf1) [anchor=north ] {\huge{.}};
 	\path (g2) ++ (-0.65,-0.38) node (nf2) [anchor=north ] {\huge{.}};
 	\node [anchor=north ] at (nf1) {$f_2$};
 	\node[anchor=north ] at (nf2) {$f_1$};

 	\path (g2) ++ (0,-0.9) node (nf0) [anchor=north ] {\huge{.}};
 	\path (g2) ++ (0,-1.1) node (f0) [anchor=north ] {$f_0$};

 	\path (g2) ++(-1.5,-2.8) node (fc1) [anchor=west, align=left] {\textit{Boundary Conditions}\\
 	$f_0 \rightarrow \vv{\bm{u}}(0,0,w)$ \\
 	$f_1 \rightarrow \vv{\bm{u}}(0,v,0)$ \\
 	$f_2 \rightarrow \vv{\bm{u}}(u,0,0)$ \\
 	$S \rightarrow$ Symmetry
 	};
 
  	\path (g2) ++ (0,2.2) node[anchor=center] {$T_0=25{ }^{\circ}$C};

	\node[anchor=center,inner sep=0] (g2) at (4,0) {\includegraphics[width = 4.4cm, trim=2cm 2.8cm 2cm 0cm]{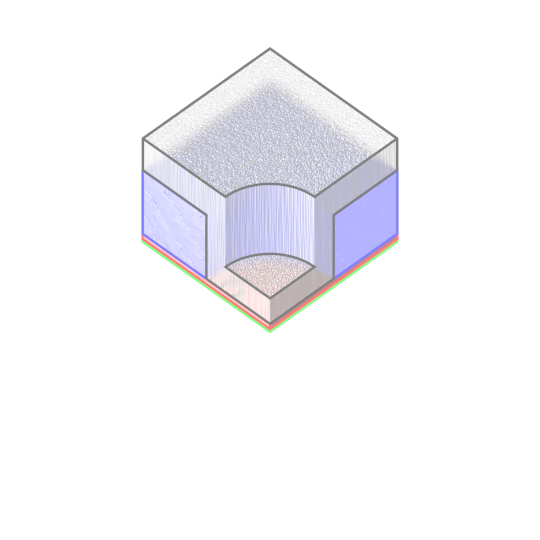}};

 	\path (g2) ++ (+1.1,-0.3) node (S1) [anchor=north ] {$S$};
 	\path (g2) ++ (-1.1,-0.3) node (S2) [anchor=north ] {$S$};
	\draw[->, anchor=center, thick, gray] (S1) -- ++ (0,0.6);
	\draw[->, anchor=center, thick, gray] (S2) -- ++ (0,0.6);
    
 	\path (g2) ++ (+0.65,-0.38) node (nf1) [anchor=north ] {\huge{.}};
 	\path (g2) ++ (-0.65,-0.38) node (nf2) [anchor=north ] {\huge{.}};
 	\node [anchor=north ] at (nf1) {$f_2$};
 	\node[anchor=north ] at (nf2) {$f_1$};

 	\path (g2) ++ (0,-0.9) node (nf0) [anchor=north ] {\huge{.}};
 	\path (g2) ++ (0,-1.1) node (f0) [anchor=north ] {$f_0$};

  	\path (g2) ++ (0,2.2) node[anchor=center] {$T_0=720{ }^{\circ}$C};

\end{tikzpicture}

	\caption{Cross-sectional diagram of the base numrical membrane model, not to scale. Boundary conditions show \CarbonC{symmetry planes to reduce the model size and} fixed constraints which restrict the velocity field ($\protect\vv{\bm{u}}$) at specific points ($\protect f$) .}
	\label{fig:layers}
\end{figure}
\RevRAO{
\subsection{Model Description}
}
The wafer stack has been modelled in the COMSOL Multiphysics\textregistered{} FEM package using the Thermal Stress module which incorporates both Solid Mechanics and Heat Transfer. A cross section schematic of the model stack is shown in Fig. \ref{fig:layers} and consists of a GaN device layer on the bottom, a lumped III-N layer, a Si frame which defines the membrane pattern and a CVD diamond layer. The lumped III-N layers represent a combination of the AlN \RevDJW{nucleation} layer and AlGaN strain relief layers, \RevGLAS{conservatively approximated as AlN}.  The base model thicknesses are given as $t_{\textrm{GaN}}=$ \SI{2}{\micro\meter}, $t_{\textrm{III-N}}=$ \SI{3}{\micro\meter}, $t_{\textrm{Si}}=$ \SI{75}{\micro\meter} and $t_{\textrm{Dia}}=$ \SI{50}{\micro\meter}. The dimensions have been chosen \RevDJW{to represent a typical MOCVD sample which has been thinned from the Si side and} cleaved to dimensions of 15 mm by 15 mm for MPCVD. \CarbonA{This sample size has been chosen since successful diamond depositions have been achieved on AlN on Si in previous work at this size in a high power density plasma\cite{Smith2020a, Mandal2019c}. Also, smaller samples minimise any diamond thickness variation over the sample owing to the small plasma size at these conditions}. The boundary conditions of the structure are also shown in Fig. \ref{fig:layers} whereby the centre point of the GaN is fixed in $xy$ space and free to displace in the $z$ direction\CarbonC{. Two symmetry planes are also defined to reduce the size of the model}.  The growth temperature of the base structure is homogenous with no gradients and chosen to match the experimental pyrometer \RevDJW{data in the centre of the sample during diamond growth for a low temperature recipe} (720 \degC{})\cite{Smith2020a}. The material parameters are the same as those used in the analytical model with additional thermomechanical properties, \RevDJW{shown in Table \ref{tab:fem}}.\cite{Sung2010,Slack1975,Roder2005,Mohr2014,Bruls2001,Shibata2007,Rounds2018}

\begin{figure}[h!]
\centering\footnotesize
    \setlength\tabcolsep{1pt}
    \begin{tabular}[t]{cccccc}
	\includegraphics[width=0.18\textwidth]{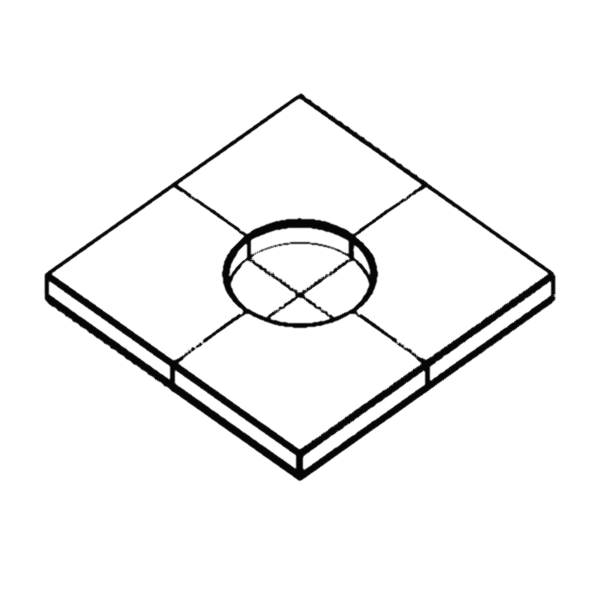} & 
	\includegraphics[width=0.18\textwidth]{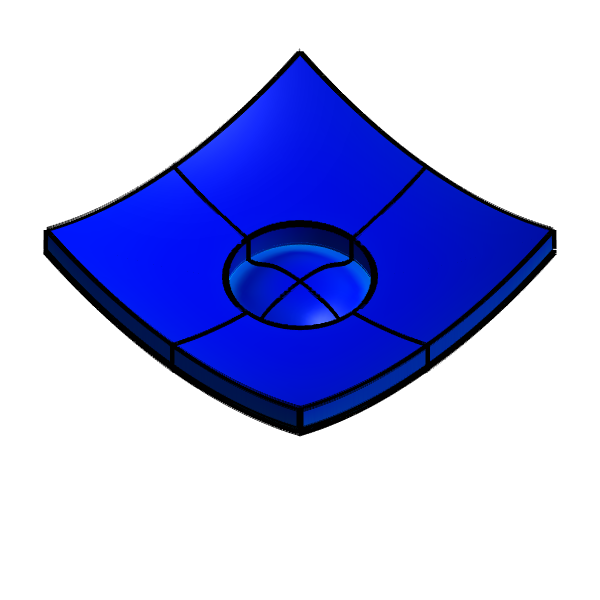} & 
	\includegraphics[width=0.18\textwidth]{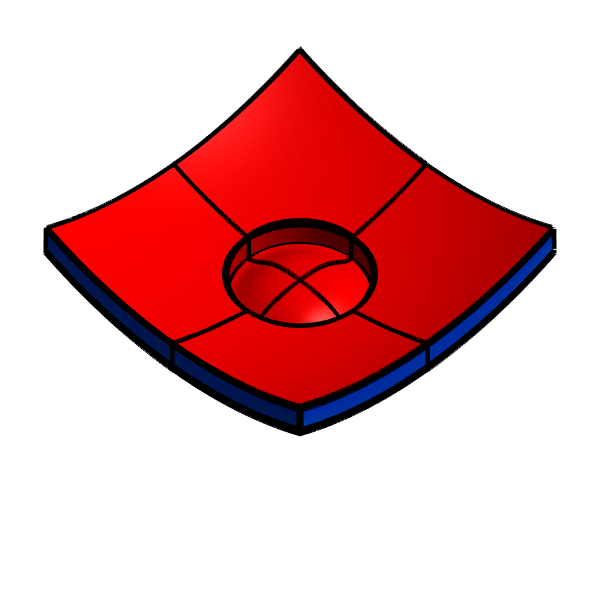}  &
	\includegraphics[width=0.18\textwidth]{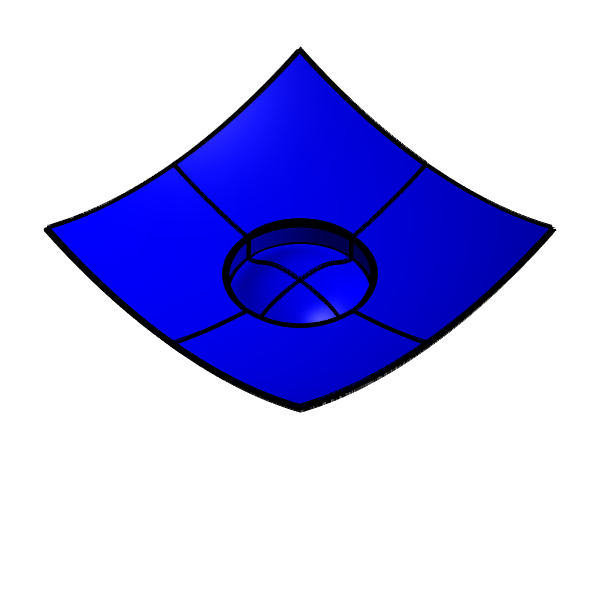} &  
	 \includegraphics[width=0.18\textwidth]{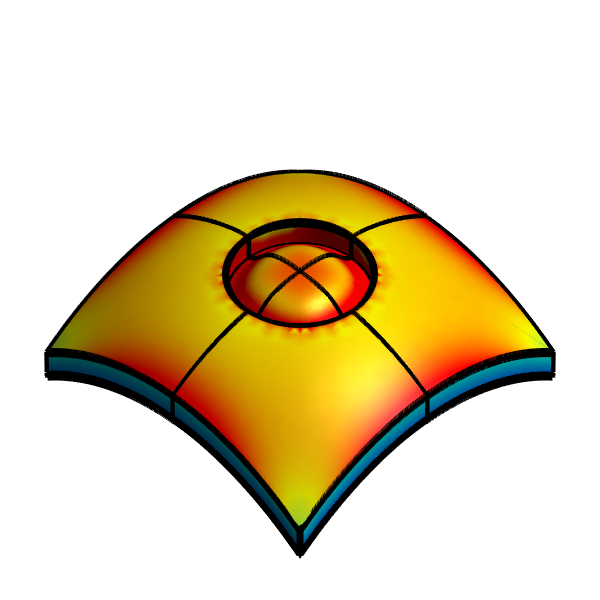} &  
 	 \includegraphics[height=0.18\textheight]{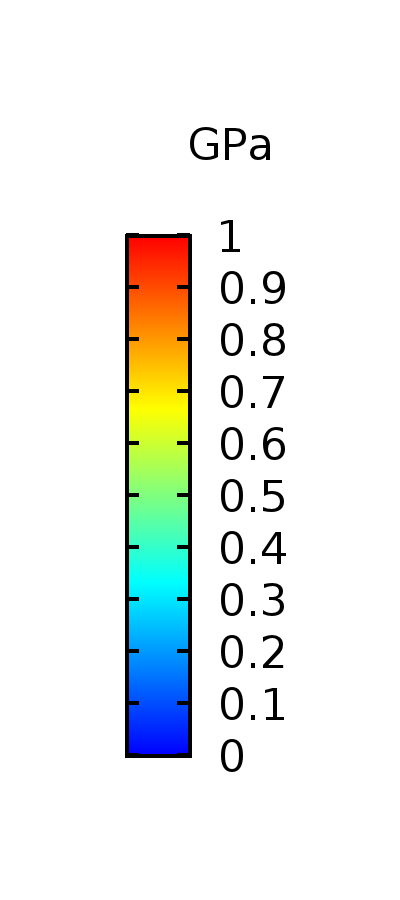}  \\
 	 0. Initial 		& 1. Membrane 		& 2. Diamond 	& 3. Diamond 			& 4. Combined cooled \\ 
	{GaN-on-Si} 	& heated to CVD	& deformed to fit 	&	stress-relieved 		& diamond and\\ 
	{membrane} 	& {temperatures} 	&{onto structure} 	&	and cooled 			& initial membrane
    \end{tabular}
  \caption{Steps taken in the FEM with colour plot showing the von Mises stress in GPa and the resultant deformation field, $\times10$ $z$ scale for visibility.}
  \label{fig:fem-ex}
\end{figure}
The model is run in an \RevCAR{automated stepped} process as is shown in Fig. \ref{fig:fem-ex}. \RevCAR{Step 0 shows the initial geometry of the model.} Step 1 is the heating model, where the structure is heated to  growth temperatures to obtain the initial thermal stresses in the absence of any diamond. Convergence of this step is crucial as it defines the initial shape of the deposited diamond. This step typically converged to two solutions, whereby the \RevGLAS{GaN-on-Si membrane} either had a positive bow (GaN displaces upwards into the Si) or a negative bow (GaN displaces away from the Si), \RevDJW{although with} the same magnitude. The presented results show the positive bow solutions since this was what was found in later experiments. Step 2 uses this displacement field to deform the diamond layer to fit on top of the stressed membrane geometry. Step 3 isolates the diamond layers and relieves the previously computed stress in this deformed configuration \RevGLAS{as it is assumed that the diamond is deposited stress-free}. The diamond layer is then cooled in isolation to obtain the deformation at room temperature. \RevBRS{The reason why the diamond is separately cooled is for stability during model convergence. This can only be assumed if there is no hysteresis in the thermal cycling}. Step 4 integrates the cooled diamond by stretching the initial room temperature membrane structure over the cooled diamond to give the final deformation and stresses associated with this state. To validate this stepped process, the absence of diamond in the final step shows no deformation.

\subsection{Numerical Model Results}
\begin{figure}[t!]
 	\setlength\tabcolsep{1pt}\centering
 	\begin{tabular}[t]{cc}
		\includegraphics[width=0.45\textwidth]{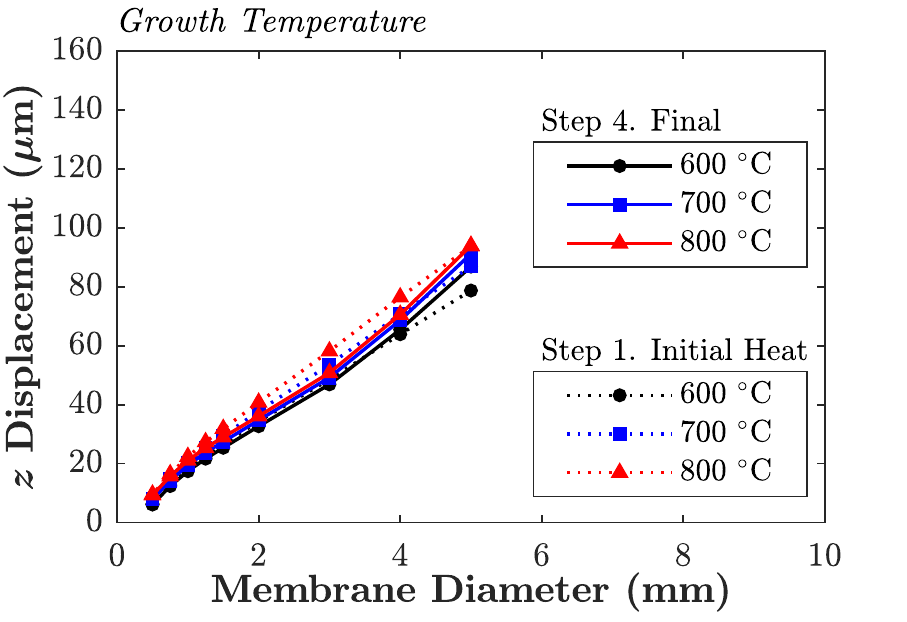} &
		\includegraphics[width=0.45\textwidth]{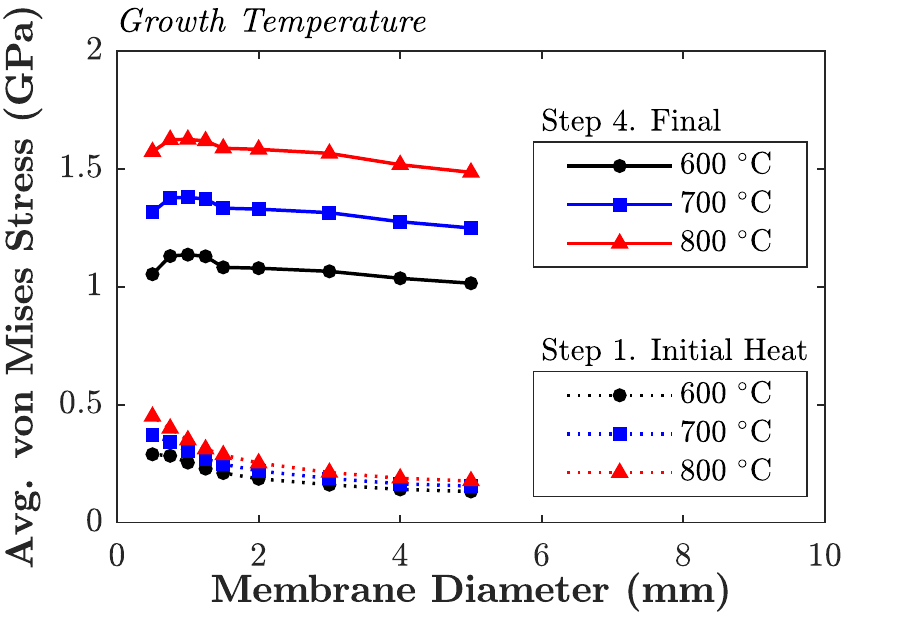} \\
		(a) & (b)
	\end{tabular}		
	\caption{Simulated results for varying the CVD growth temperature are given in \CarbonB{(a) and (b) for the membrane bow and the average residual stress in the GaN membrane, relative to step 0, respectively}. Base model layer thicknesses have been used: $t_{\textrm{GaN}}=$ \SI{2}{\micro\meter}, $t_{\textrm{III-N}}=$ \SI{3}{\micro\meter}, $t_{\textrm{Si}}=$ \SI{75}{\micro\meter} and $t_{\textrm{Dia}}=$ \SI{50}{\micro\meter}. }
  \label{fig:fem-T}
\end{figure}

\begin{figure}[t!]
 	\setlength\tabcolsep{1pt}\centering
 	\begin{tabular}[t]{cc}
		\includegraphics[width=0.45\textwidth]{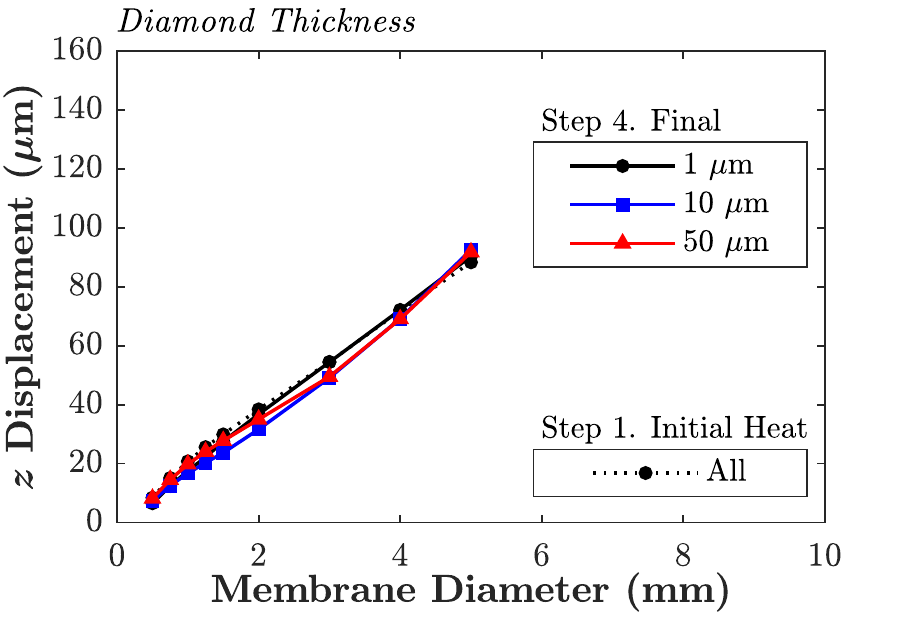} &
		\includegraphics[width=0.45\textwidth]{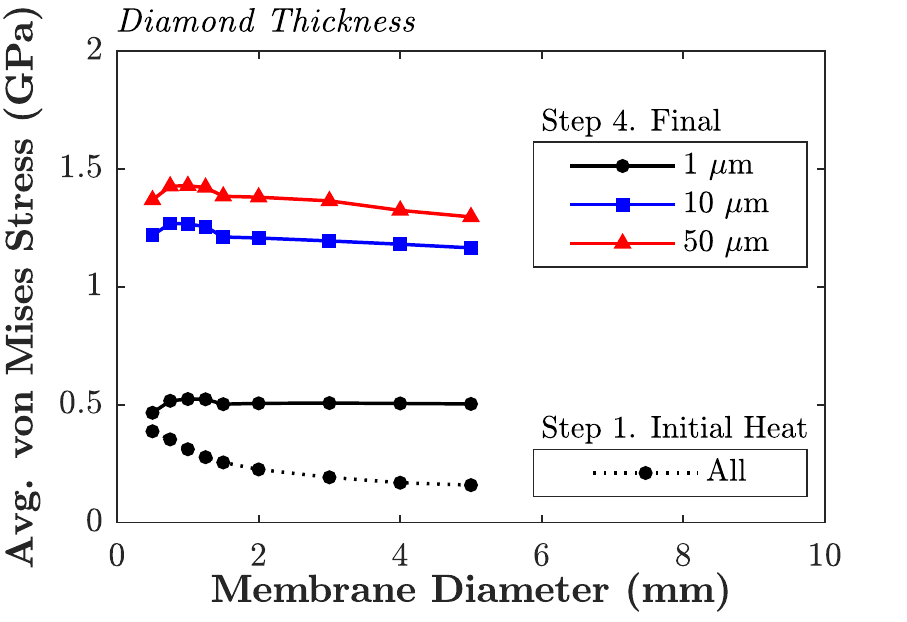} \\
		(a) & (b) \\ {} & {} \\
		\includegraphics[width=0.45\textwidth]{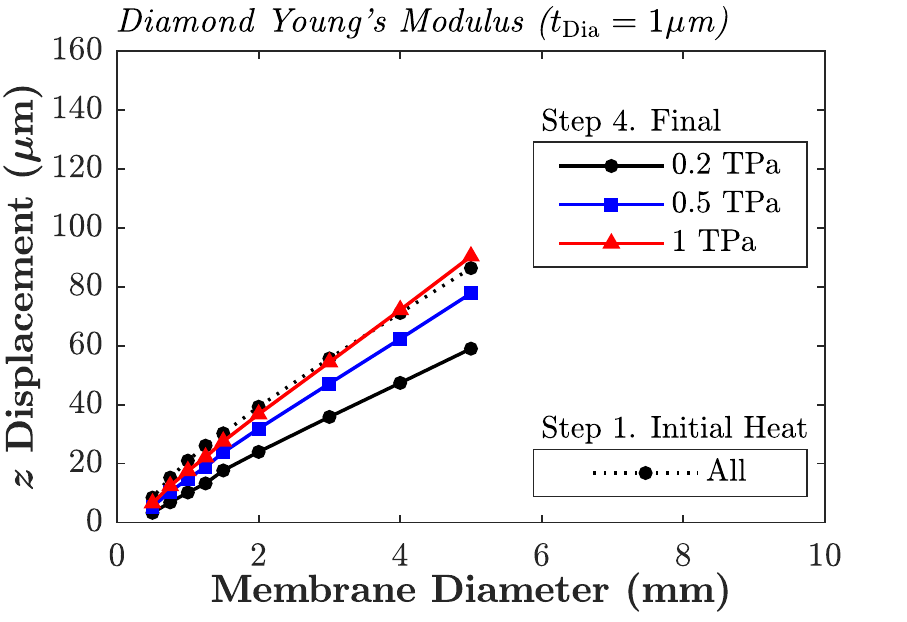} &
		\includegraphics[width=0.45\textwidth]{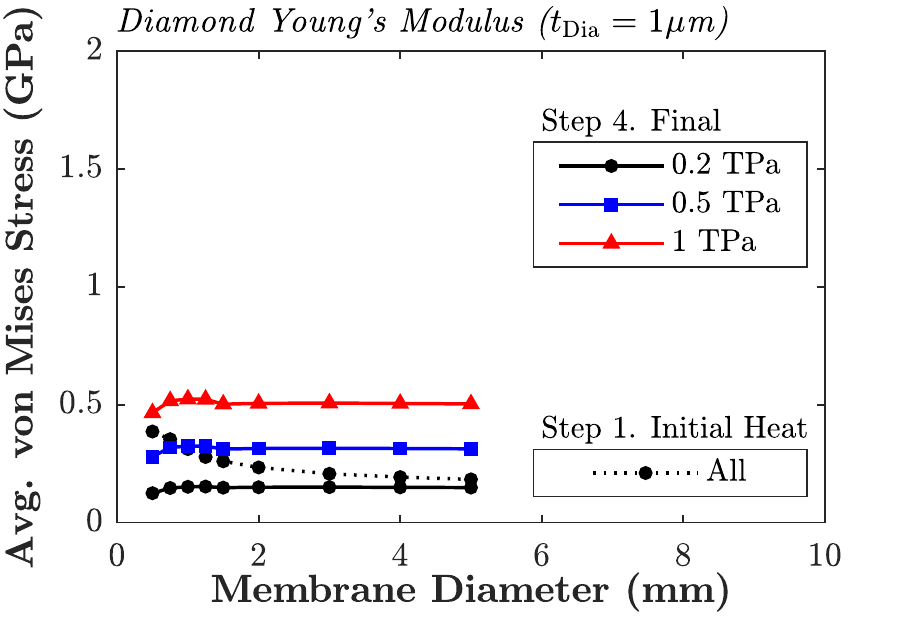} \\
		(c) & (d)
	\end{tabular}		
	\caption{Simulated results for varying $t_{\textrm{Dia}}$ are given in \CarbonB{(a) and (b) for the membrane bow and the average residual stress in the GaN membrane relative to step 0, respectively}. Results for varying $E_{\textrm{Dia}}$ are given in (c) and (d) for the membrane bow and stress, respectively. The traces marked `All' refer to the heating phase of all the model cases since the diamond is introduced afterwards. The base model layer thicknesses have been used for other layers: $t_{\textrm{GaN}}=$ \SI{2}{\micro\meter}, $t_{\textrm{III-N}}=$ \SI{3}{\micro\meter} and $t_{\textrm{Si}}=$ \SI{75}{\micro\meter}. A deposition temperature of $T_g=720$ \degC{} was used as to simulate low temperature growth recipes for GaN/III-N membranes\cite{Smith2020a}. The base value of $E_{\textrm{Dia}} = 1$ TPa.}
  \label{fig:fem-dia}
\end{figure}

\begin{figure}[t!]
 	\setlength\tabcolsep{1pt}\centering
 	\begin{tabular}[t]{cccc}
		\includegraphics[width=0.45\textwidth]{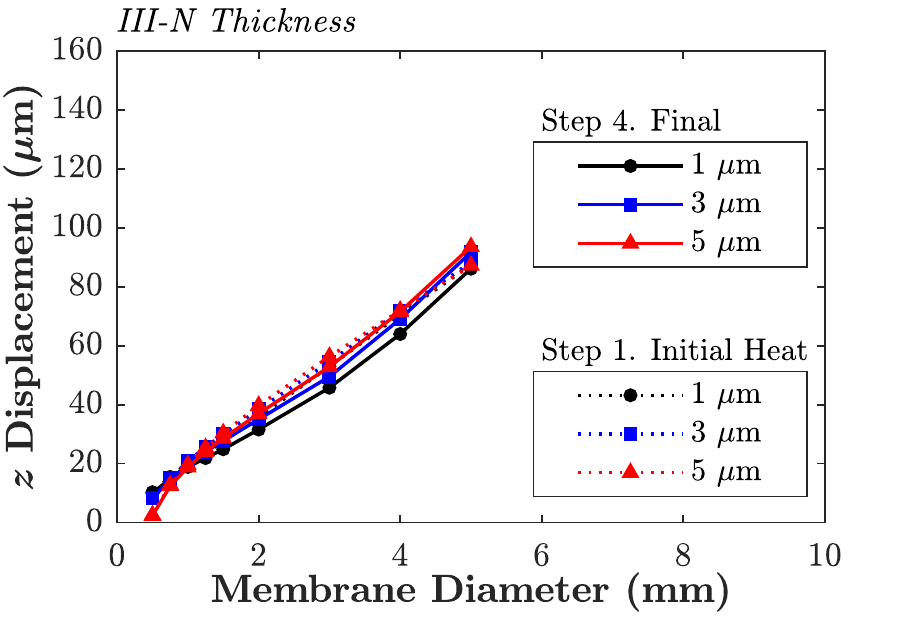} &
		\includegraphics[width=0.45\textwidth]{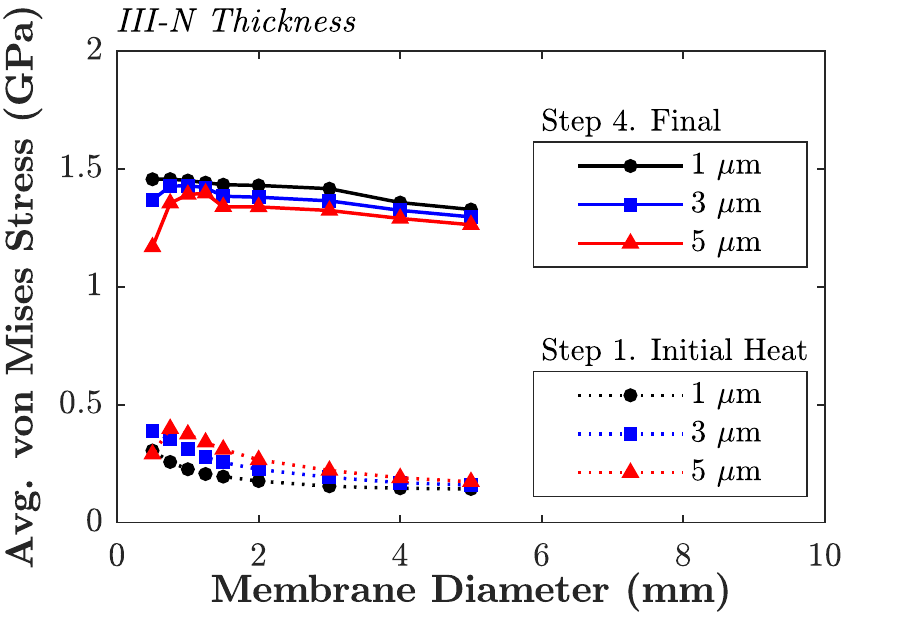} \\
		(a) & (b) \\ {} & {} \\
		\includegraphics[width=0.45\textwidth]{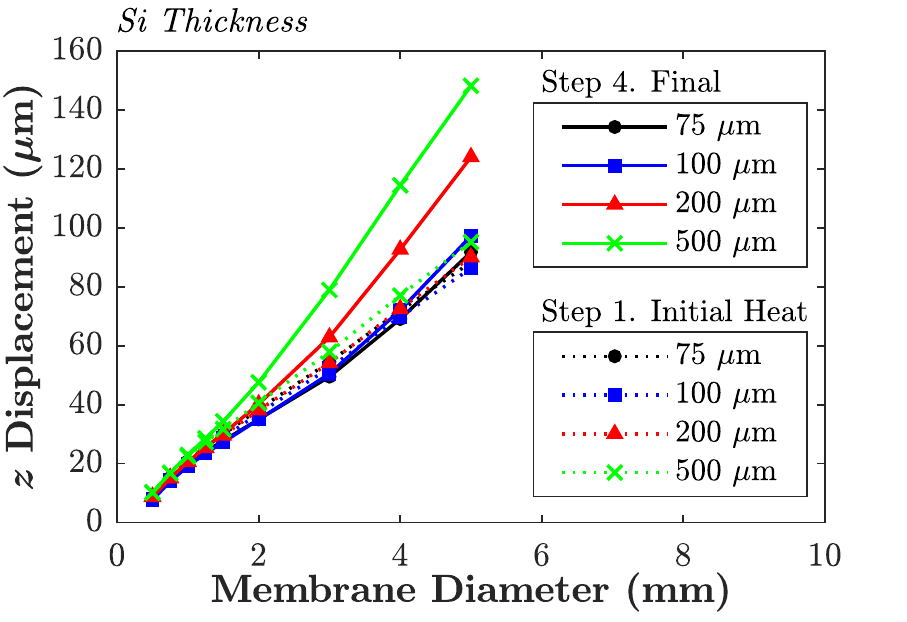} &
		\includegraphics[width=0.45\textwidth]{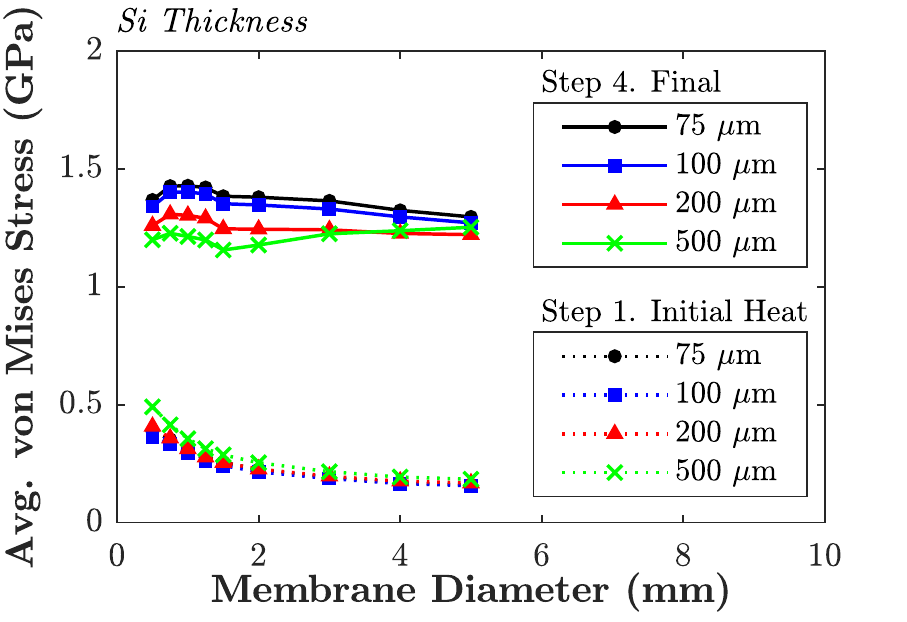} \\
		(c) & (d) \\
	\end{tabular}		
	\caption{Simulated results for varying $t_{\textrm{III-N}}$ are given in \CarbonB{(a) and (b) for the membrane bow and the average residual stress in the GaN membrane, relative to step 0, respectively}. Results for varying $t_{\textrm{Si}}$ are given in (c) and (d) for the membrane bow and stress, respectively. The base model layer thicknesses have been used for the other layers: $t_{\textrm{GaN}}=$ \SI{2}{\micro\meter}, $t_{\textrm{III-N}}=$ \SI{3}{\micro\meter}, $t_{\textrm{Dia}}=$ \SI{50}{\micro\meter} and $t_{\textrm{Si}}=$ \SI{75}{\micro\meter}. A deposition temperature of $T_g=720$ \degC{} was used as to simulate low temperature growth recipes for GaN/III-N membranes\cite{Smith2020a}}
  \label{fig:fem-alnsi}
\end{figure}

The modelled membrane results are shown in Fig. \ref{fig:fem-T}, \ref{fig:fem-dia} and \ref{fig:fem-alnsi} where the dotted lines show the heated sample and the solid lines show the final state where the cooled diamond is integrated with the structure. A positive $z$ displacement refers to a membrane bow into the Si while a negative $z$ displacement refers to away from the Si, as per the defined coordinate system in Fig. \ref{fig:layers}. The von Mises stress denote the residual stress associated with the displacement fields.


Fig. \ref{fig:fem-T} shows the base model \CarbonB{displacement as a function of membrane diameter and at varying growth temperatures}. It is clear from Fig. \ref{fig:fem-T}.a that the bow increases with membrane diameter. \RevRAO{It \CarbonC{is} also shown that, as expected, the bow increases with deposition temperature, although to a lesser degree, resulting in an increase in stress}. The heated stresses in Fig. \ref{fig:fem-T}.b are approximately 0.2 GPa for larger membranes and increase to 0.4 GPa for smaller membranes. These values are in the same range as the heated stresses for the analytical model. \CarbonC{However,} the heated bow is much larger than the analytical model; for example in Fig. \ref{fig:fem-T} upon changing the temperature from 600 to 800 \degC{}, the 5 mm membrane bows\RevDJW{, respectively,} from 80 to \SI{90}{\micro\metre} in the numerical model while in Fig. \ref{fig:ana}, the calculated bow for the \CarbonC{same membrane is 22 to \SI{30}{\micro\metre} in the analytical model}. \CarbonC{This large  difference} is attributed to the additional \CarbonC{thermal stress contribution} from the Si border which \CarbonC{is} not taken into account in the analytical approach, demonstrating a significant limitation in using \RevDJW{Eq.} (\ref{eq:thermalstress}) with (\ref{eq:stoney}) to predict the bow in membrane structures. After cooling with the diamond introduced, the overall bow is relatively unchanged, implying that when the diamond is deposited, it attempts to hold the deformed membrane into place upon cooling owing to its extreme hardness and low CTE across the temperature range. \RevRAO{This results in a large increase} in the residual stresses, as the \RevRAO{initial membrane model in Step 0 should be flat at room temperature} but is held in the bowed state. These residual stresses also decrease \CarbonC{as the growth temperature is reduced}. \RevRAO{As shown in Fig. \ref{fig:fem-T}.b, GaN/III-N membranes with diameters between 1 and 5 mm will be under a stress between 1.0 to 1.7 GPa which is still less than the tensile strength (4 to 7.5 GPa at room temperature)\cite{Brown2011}}.

Fig. \ref{fig:fem-dia} shows the models for varying the diamond thickness, where the traces marked `All' stipulate the result for all cases for the heating models; the initial geometry is not varying only the diamond and its properties. A lower limit of \SI{1}{\micro\metre} has been chosen since it has been shown that thinner diamond films result in incredibly low thermal conductivities\cite{Zhou2017}. The diamond thickness appears to have minimal effect on the observed bow but a significant effect on the residual stress in the membrane. This result also demonstrates that even a \SI{1}{\micro\metre} thin diamond layer is enough to hold the  membranes into the heated bowed state by several tens of  micrometres owing to the extreme stiffness of diamond. To further demonstrate this, Fig. \RevGLAS{\ref{fig:fem-dia}.c and \ref{fig:fem-dia}.d show the results for varying the diamond Young's Modulus ($E_{\textrm{Dia}}$)} for a \SI{1}{\micro\metre} thick diamond layer. Decreasing the stiffness decreases the final membrane bow and stress. At very low values of around 200 GPa, the diamond is no longer rigid enough to fully sustain the deformation thereby reducing the final bow and stress. However, such low values of  $E_{\textrm{Dia}}$ are not expected for CVD diamond, even for ultra-nano-crystalline (UNCD) films\cite{Mohr2014,Williams2010}. 
 
Fig. \ref{fig:fem-alnsi}.a and \ref{fig:fem-alnsi}.b show the effect of varying the III-N stack thickness, revealing minimal differences in both bow and cooled stress. The heated \CarbonC{bow and} stress, however, increases with thickness which is not favourable at high temperatures since GaN weakens towards 1000 \degC{}\cite{Yonenaga2001}. \RevGLAS{A similar case is found} for increasing the Si thickness\CarbonC{, although a much larger bow is seen in Fig.\ref{fig:fem-alnsi}} across all membrane sizes. The average stress in the membrane is also marginally relieved upon cooling as the Si is thickened, shown in Fig. \ref{fig:fem-alnsi}.d. Since the Si thickness can be varied by a much larger range at tens of microns compared to the III-N layers at a few microns, changes to the Si are expected to have a more profound effect on the bow and stress than the III-N layers. \RevDJW{This implies a potential route to minimising the bow through Si thickness, however it would need to be much less than \SI{75}{\micro\metre}}.

In the presented models, it is apparent that after CVD diamond deposition, \RevCAR{the final stress in the membranes is within the mechanical limits of GaN, however, the anticipated bow of large membranes is  significant, imposing challenges for device manufacturing through contact photo-lithography}. \RevBRS{The models imply that only small membranes are a potentially viable route. To verify these findings, circular GaN/III-N membrane samples of varying diameter were fabricated with similar dimensions to the models. The effect of CVD diamond deposition on the mechanical properties was investigated through measurement.}

 \section{Experiment}\label{sec:exp}
 
\CarbonA{There were 3 samples in total, fabricated using commercially obtained  wafers}, diced to square dimensions of 15 mm by 15 mm. The fabrication of the membranes and details can be found in previous work\cite{Smith2020a}. In brief, the as received GaN/III-N on Si wafer was etched from the Si side using a high power ICP at 900 W, followed by membrane patterning using photolithography and finally a lower power ICP etch at 600 W to completely remove the Si and expose an AlN layer on the GaN/III-N membranes. The thickness of the remaining Si handle border was was measured using a micrometer. Following this, CVD diamond growth was achieved using a Carat systems CTS6U MPCVD reactor. \CarbonA{The temperature of the sample was measured using a Williamson pyrometer (Model DWF-24-36C) which uses dual wavelengths at around \SI{2}{\micro\metre} to minimise emissivity errors}. Prior to growth, a non-ultrasonic seeding method was used owing to the fragility of the membranes. Briefly, the exposed GaN/III-N membranes were first pre-treated in a \nitro{}/\hydro{} microwave plasma with a forward power of 1.5 kW at 20 Torr. \CarbonA{This step allows some control over the seeding density by decreasing the streaming potential of the AlN. This brief plasma pre-treatment stage has been shown to increase the oxygen surface content on AlN  by partially etching and subsequent adsorption of oxygen onto the surface once the sample is brought to atmosphere\cite{Mandal2019c}}. Following this, a nano-diamond colloid solution was pipetted onto the exposed GaN/III-N membrane side, covering both the membrane and the Si handle border. The samples were rinsed carefully in de-ionised water and dried on a hotplate for 10 minutes set to 115\degC{}. Diamond growth was achieved in a \meth{}/\hydro{} microwave plasma with a forward power of 5.5 kW at 110 to 120 Torr and a \meth{} concentration of 3\% at a total flow rate of 500 sccm. At these conditions, the pyrometer measured a temperature of around 720 to 750 \degC{}. With a typical growth rate at these conditions of approximately $2$ to \SI{3}{\micro\metre}/hour, growth runs were conducted for 19 hours to achieve a diamond thickness of around 38 to \SI{57}{\micro\metre}.

\begin{figure}[t!]
	\centering
	\footnotesize
	\setlength\tabcolsep{1pt}
	\begin{tabular}[t]{ccc}
		\multicolumn{1}{l}{\textbf{Before}} &{} &{} \\
		\includegraphics[width=0.18\textwidth]{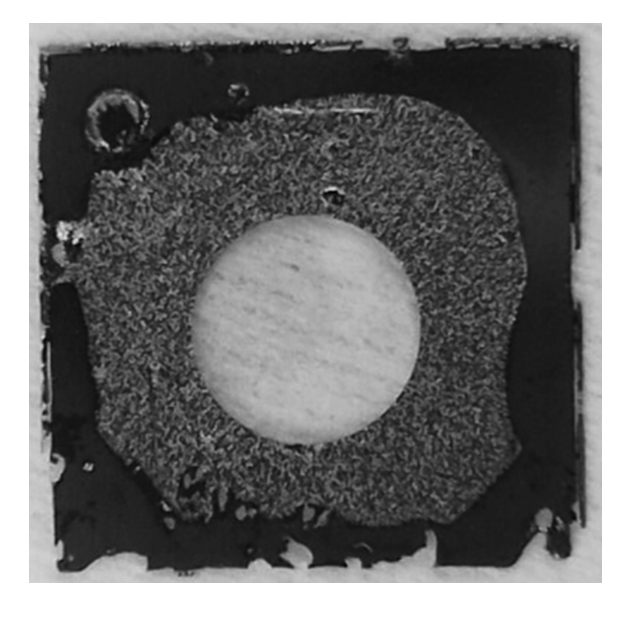} & 
		\includegraphics[width=0.18\textwidth]{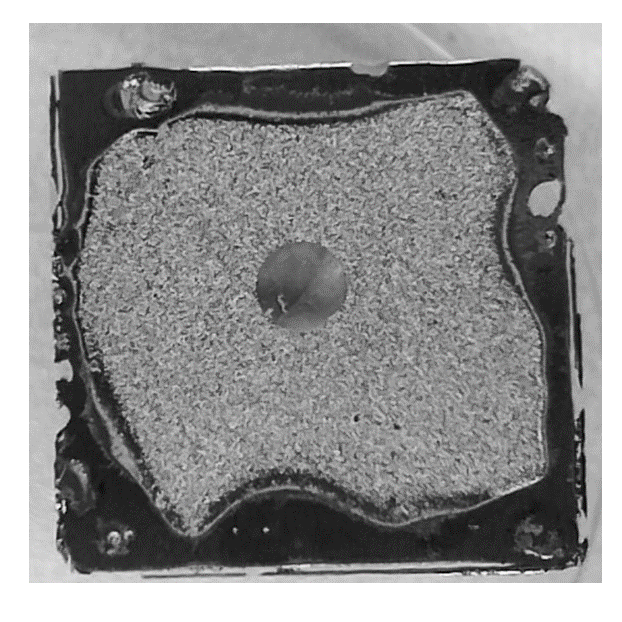} & 
		\begin{tikzpicture}
			\node[anchor=center,inner sep=0] at (0,0)
			{\includegraphics[width=0.18\textwidth]{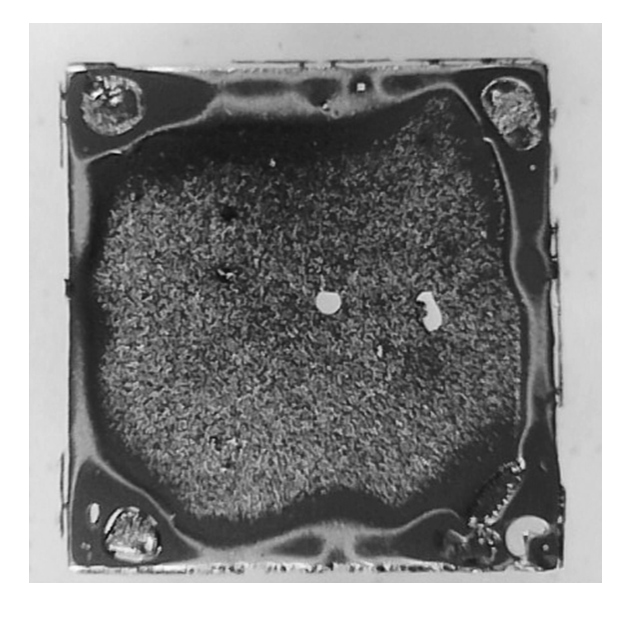}};
			\draw[->, anchor=east, very thick, yellow] (0.45,0.4) -- ++ (-0.3,-0.3);
		\end{tikzpicture}
		\\
	\end{tabular} \\
	
	\begin{tabular}[t]{ccc}
		\multicolumn{1}{l}{\textbf{After}} &{} &{} \\
		\includegraphics[width=0.18\textwidth]{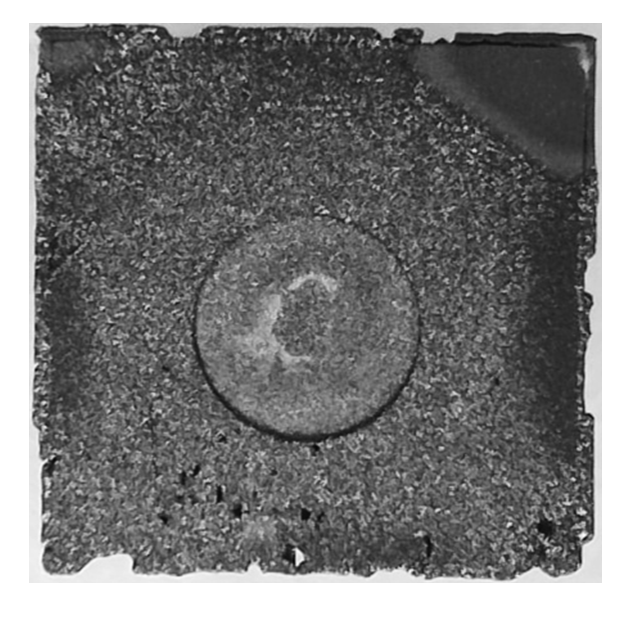} & 
		\includegraphics[width=0.18\textwidth]{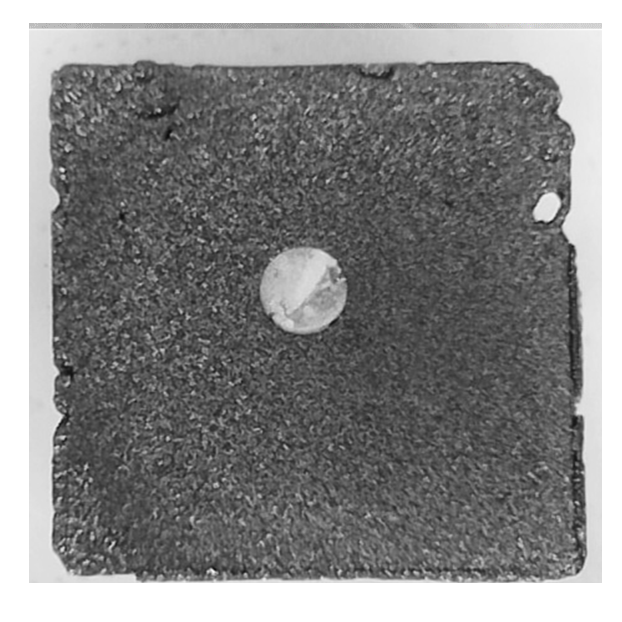} & 
		\begin{tikzpicture}
			\node[anchor=center,inner sep=0] at (0,0)
			{\includegraphics[width=0.18\textwidth]{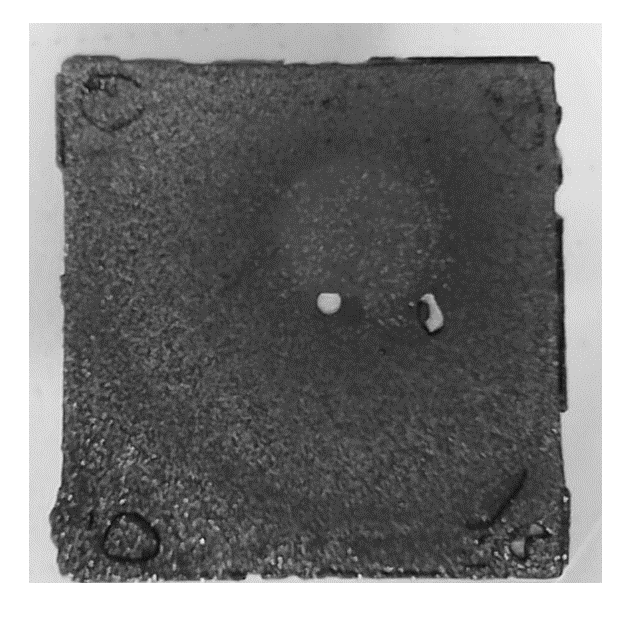}};
			\draw[->, anchor=east, very thick, yellow] (0.45,0.4) -- ++ (-0.3,-0.3);
		\end{tikzpicture} 
		\\
		5 mm &  2 mm & 0.5 mm
	\end{tabular} \\
	\begin{tabular}[t]{cl}
		\multicolumn{1}{l}{\textbf{Initial Heating}} &{} \\
		\includegraphics[width=0.27\textwidth]{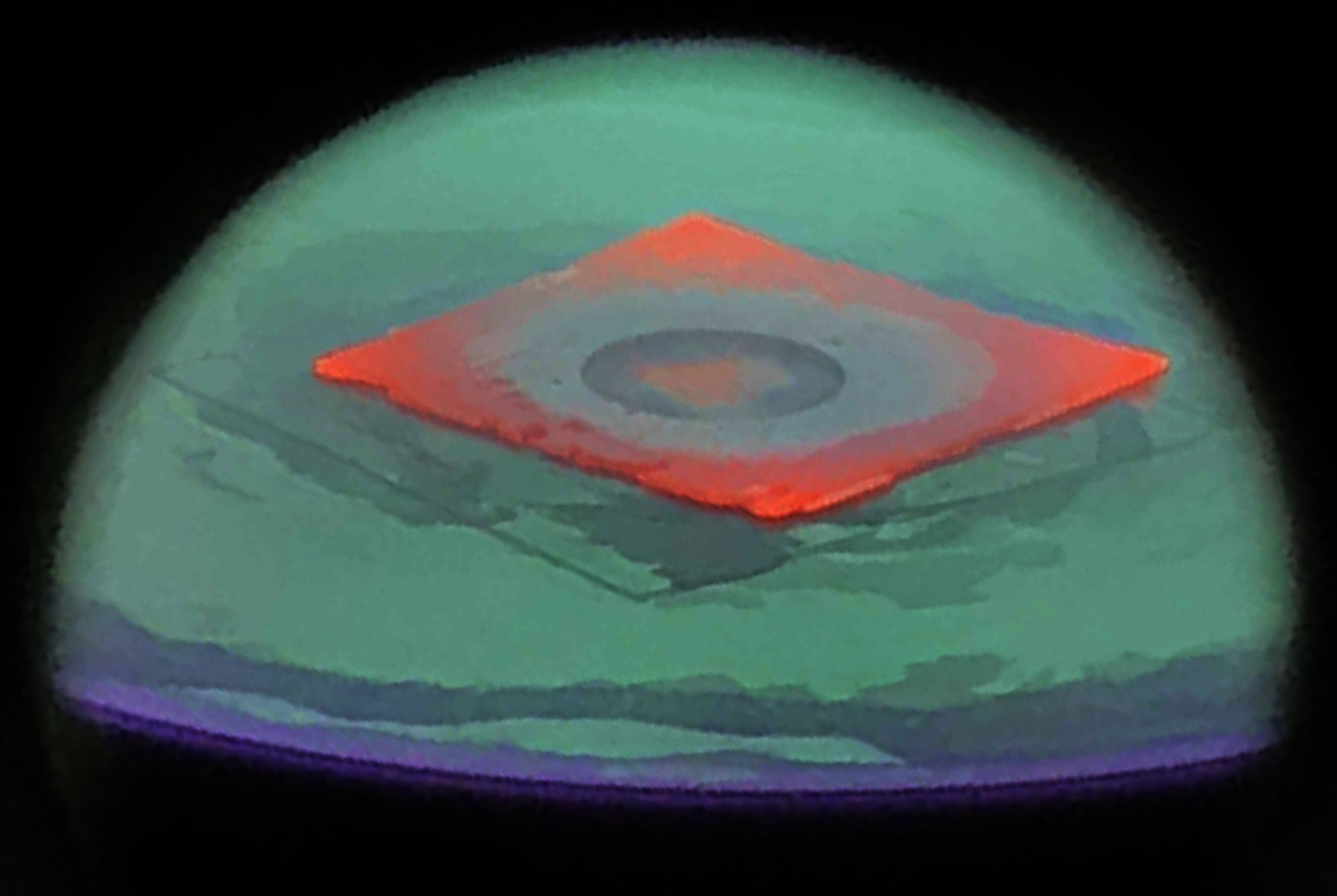} &
		\includegraphics[height=0.125\textheight]{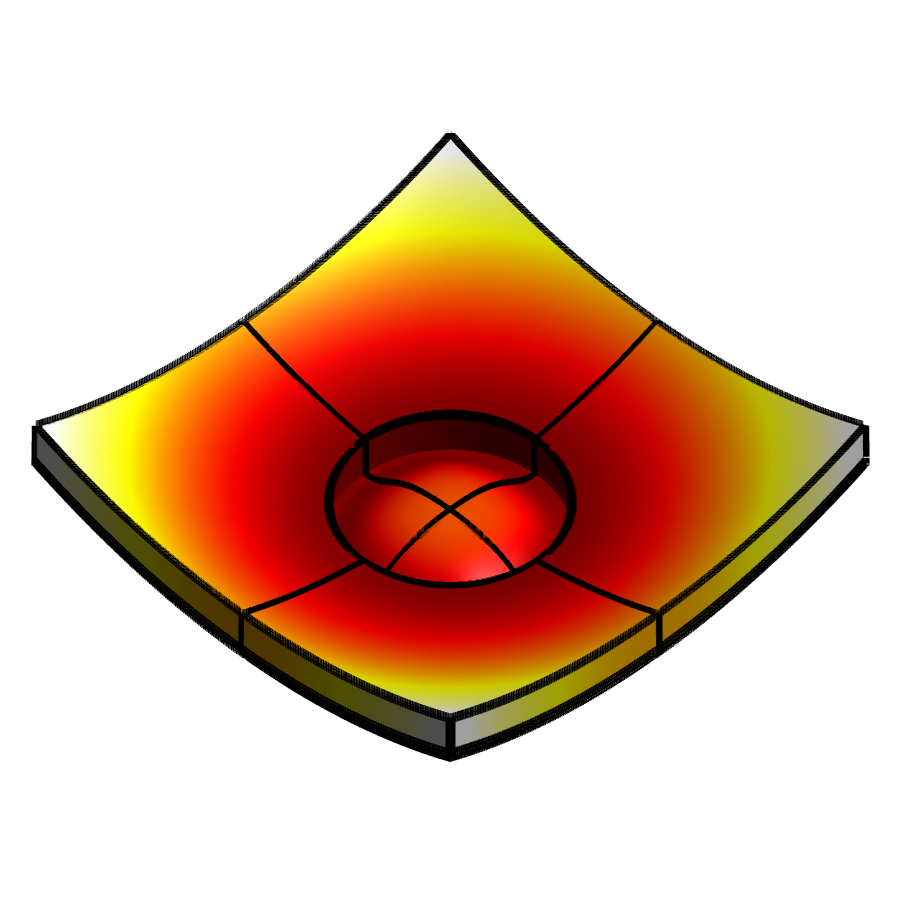}
		\includegraphics[height=0.125\textheight]{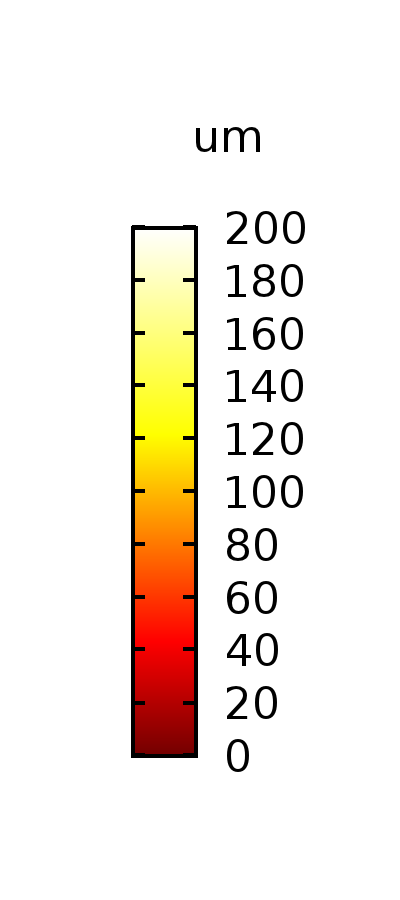}
	\end{tabular}
	\caption{\CarbonA{Photos of the 3 GaN/III-N membrane samples} before (top row), after MPCVD diamond growth (middle row) and at the beginning of MPCVD growth (bottom row) with numerical model of the displacement field in micrometres ($z$ scale $\times$10 for visibility). All samples are grown one by one, positioned centrally on the sample holder as shown in the bottom photo.
	} 
	\label{fig:photo}
\end{figure}

Following growth, surface profilometry was used to determine the membrane bow with a profiler tip of diameter of \SI{12}{\micro\metre}, a force of 1 mg and a maximum travel time of 60 seconds across the membrane. Raman spectroscopy was used to determine the \CarbonC{structural signatures of the} GaN/III-N and CVD diamond and infer the stress in the membranes, achieved using a Renishaw inVia Raman spectrometer with a 514 nm laser. To ensure that there is an adhered interface between the CVD diamond and the N-polar side \CarbonC{of the} III-N stack, \RevRAO{cross-sectional observation of the samples was conducted by high angle annular dark field scanning transmission electron microscopy (HAADF-STEM) and energy dispersive X-ray spectroscopy (EDX) in an FEI Tecnai Osiris microscope operated at 200 kV. The TEM foil was prepared by focused ion beam (FIB) from the \CarbonC{GaN/}III-N side of the membrane}.  

The fabricated membranes are shown in Fig. \ref{fig:photo}, before, after and at the start of CVD diamond growth. After deposition, the 0.5 and 2 mm \RevDJW{diameter} membranes have high optical transparency, implicit of a successful growth, however, the 5 mm diameter membrane shows a much darker film implying \CarbonA{that the GaN membrane is significantly damaged}. This is likely due to the membrane bowing upwards towards the \meth{}/\hydro{} microwave plasma \CarbonA{and melting}. Referring to the image of the sample inside the MPCVD chamber during the first hour of growth in Fig. \ref{fig:photo}, it is clear that the \CarbonB{centre of the membrane is at elevated temperatures. Also, the sample corners are at much higher temperatures than the rest of the sample}. \RevRAO{This is corroborated by the displacement fields of the numerical model that shows that these regions are significantly elevated during growth}.

\begin{figure}[t]\centering
	\includegraphics[width=0.8\textwidth, trim= {0 2cm 0 2cm}]{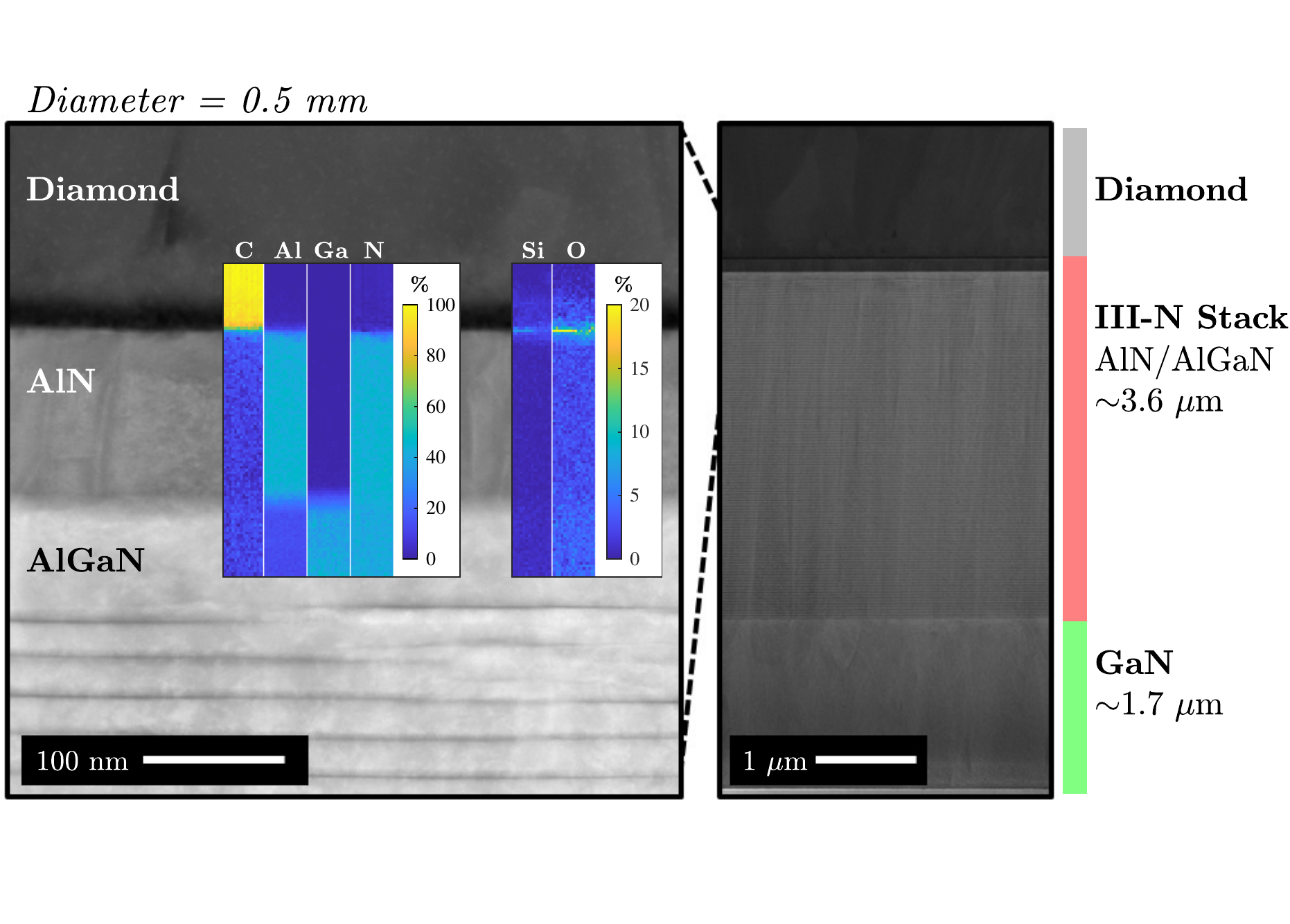}
	\caption{\CarbonA{Cross-sectional \RevRAO{HAADF-STEM} microgaph and EDX data of CVD diamond grown on a 0.5 mm GaN/III-N membrane.}} 
	\label{fig:stem}
\end{figure}

\subsection{Scanning Transmission Electron Microscopy}

\CarbonA{A Cross sectional STEM micrograph} of \CarbonC{the CVD diamond grown on the GaN/III-N membrane stack (0.5 mm diameter)} is shown in Fig. \ref{fig:stem}. The EDX analysis shows the carbon rich layer bound to an aluminium and nitrogen rich layer, where the former is the deposited CVD diamond layer and the latter corresponds to an AlN \RevDJW{nucleation} layer ($\sim$\SI{130}{\nano\metre}). Underneath the AlN \RevDJW{nucleation} layer, \RevRAO{the AlGaN strain relief layers can be clearly seen} ($\sim$\SI{3.6}{\micro\metre}) which is then followed by the GaN device layer ($\sim$\SI{1.7}{\micro\metre}) and \RevDJW{AlGaN barrier layers which forms the device 2DEG}. In between the diamond and the AlN \RevDJW{nucleation} layer shows \CarbonA{a very thin layer of silicon and oxygen, most likely SiO$_2$ based on the atomic ratios. This is a curious finding owing to the absence of any oxygen in the gas precursors. There are two possible origins for this SiO$_2$ layer. The first is from leftover Si after the ICP etch which then forms an oxide when brought to atmosphere the MPCVD stages. The second is that the Si border is etched in the MPCVD plasma pre-treating stage before seeding, resulting in redeposition of the Si on the membrane and subsequent oxide formation when the sample is brought to atmosphere; more discussion on this is given later. }

\subsection{Raman Spectroscopy}

\begin{figure}[t!]
	\centering
	\includegraphics[width=0.6875\textwidth]{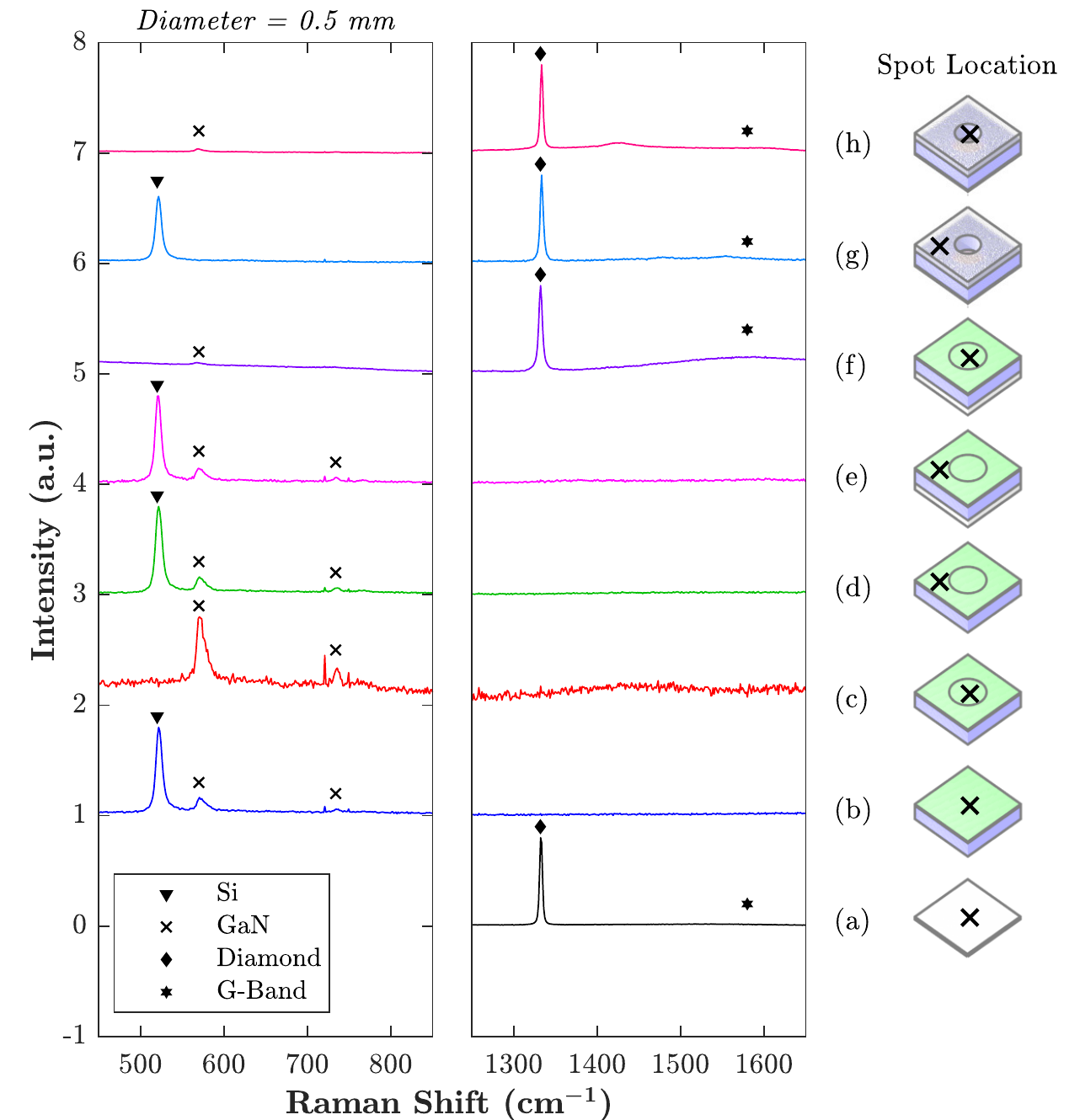}
	\caption{\RevBRS{Raman spectra of the 0.5 mm GaN membrane sample at stages during the fabrication process. Spectra were taken at different spot locations for (a) free-standing diamond reference, (b) as received GaN-On-Si wafer, (c) and (d) GaN side after membrane fabrication, (e) and (f) GaN side after CVD diamond growth and (g) and (h) diamond side after CVD diamond growth.} Background subtraction has been achieved using a linear fit.}
	\label{fig:raman}
\end{figure}

\begin{figure}[h!]
	\centering
	 \begin{tabular}[t]{cc}
		\includegraphics[width=0.45\textwidth]{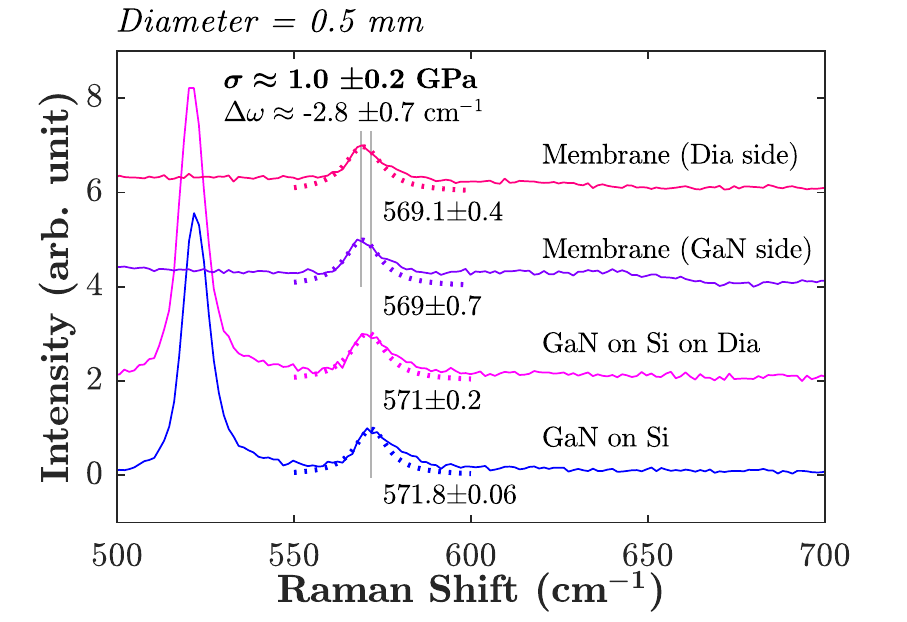} &
		\includegraphics[width=0.45\textwidth]{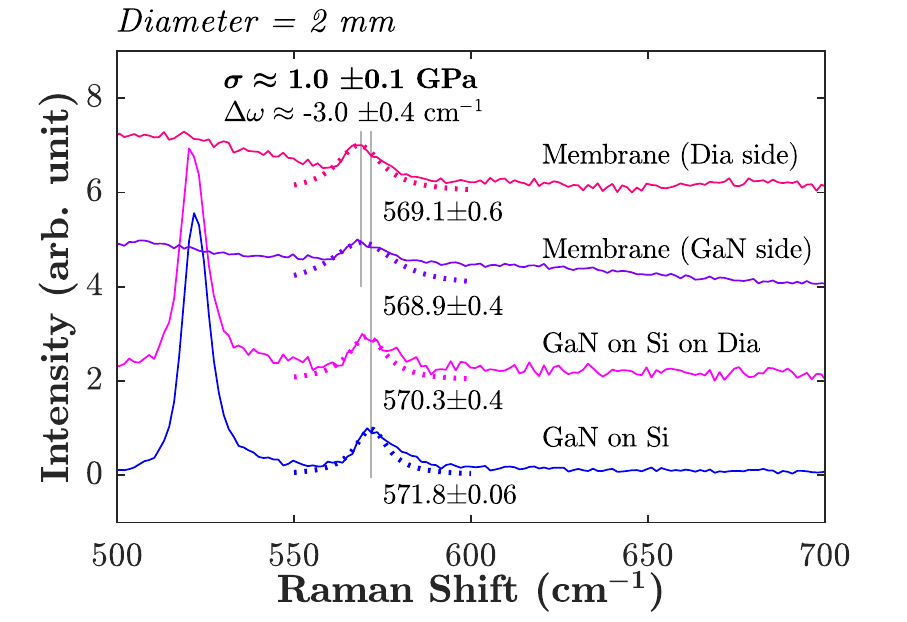} \\
		(a) & (b) \\
		\includegraphics[width=0.45\textwidth]{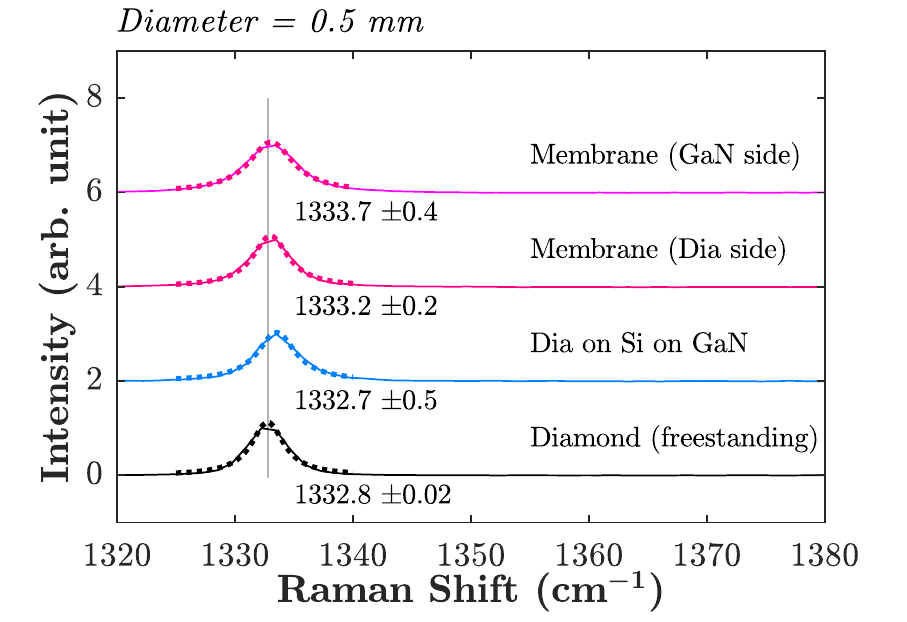} &
		\includegraphics[width=0.45\textwidth]{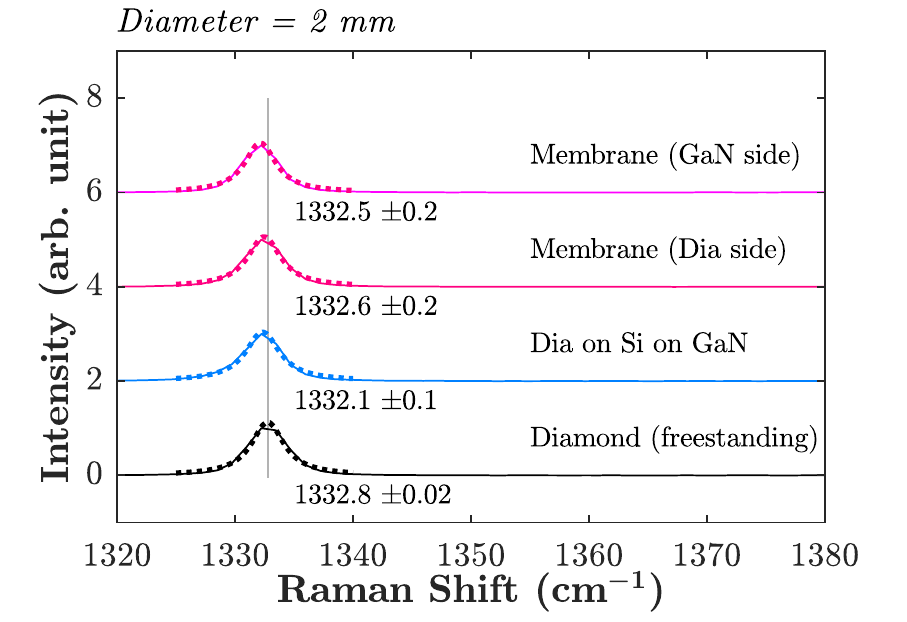} \\
		(c) & (d)
	\end{tabular}		
	\caption{Magnified view of the Raman spectra of the GaN/III-N membranes. (a) and (b) show the spectra at the Si peak and GaN $E_2$(high) peak while (c) and (d) show the spectra at the diamond peak for membrane diameters of 0.5 mm and 2 mm, respectively. Color schemes are the same as Fig. \ref{fig:raman}. Numerical values are the peak average with standard deviation of 3 different measurements of a similar region. Spectra obtained with a laser wavelength of 514 nm and background subtraction using a linear fit while Lorentzian peak fits are obtained using least squares curve fitting. }
	\label{fig:raman-zoom}
\end{figure}

The Raman spectra of the 0.5 mm membrane sample is shown in Fig. \ref{fig:raman} at different stages of the fabrication process. Fig. \ref{fig:raman}.a shows a free-standing diamond sample produced at similar growth conditions with \CarbonC{the} Si removed. This sample shows a sharp peak at 1333 \cm{} associated with diamond and minimal signs of non-diamond carbon impurities. Fig. \ref{fig:raman}.b shows the as-received GaN-on-Si wafer with the Si substrate contribution at 520 \cm{} and the GaN/III-N layers at around 569 \cm{} and 734 \cm{}, corresponding to the $E_2$(high) and the $A_1$ longitudinal optical modes of GaN, respectively\cite{Kuball2001}. After photolithography and ICP etching to create the membrane, Fig. \ref{fig:raman}.c shows the spectra on the membrane and demonstrates that no Si mode could be detected, implicit of a successful etch, while Fig. \ref{fig:raman}.\CarbonC{d} shows that on the Si \RevDJW{border} the GaN/III-N features are relatively unchanged. Fig. \ref{fig:raman}.e to Fig. \ref{fig:raman}.h show the spectra after diamond growth. Fig. \ref{fig:raman}.e shows that after growth the GaN/III-N on the Si border is minimally changed, however, on the membrane in Fig. \ref{fig:raman}.f, there is also a sharp diamond peak at around 1333 \cm{}. This is due to the $\sim$\SI{50}{\micro\metre} thick diamond layer on the underside of the GaN/III-N membrane. There also exists very small and broad D and G bands at 1350 \cm{} and 1580 \cm{}, attributed to a very low concentration of non-diamond (graphitic like) carbon that is introduced during MPCVD growth. \RevCAR{Fig. \ref{fig:raman}.g shows the diamond grown on the Si}. Finally, Fig. \ref{fig:raman}.h shows the Raman spectra on the GaN membrane from the diamond side which has a similar spectra to the membrane side in Fig \ref{fig:raman}.f.

Since the diamond layer is much thicker than the GaN/III-N layers, the high intensity diamond peak dominates the spectra on the membrane, masking the GaN/III-N features. Magnified views of the GaN and diamond regions are given in Fig. \ref{fig:raman-zoom}. For both the 0.5 mm and 2 mm samples, the GaN $E_2$(high) peaks are clearly shown, however, for the 5 mm sample this peak was not consistently found at multiple regions on the membrane. This is likely caused by damage owing to the elevated regions as is shown in Fig. \ref{fig:photo}. For the 0.5 mm and 2 mm samples\CarbonC{, the} GaN $E_2$(high) peaks are shifted to lower frequencies of $-2.8\pm0.7$ \cm{} and $-3.0\pm0.4$ \cm{}, respectively. This implies that the GaN/III-N layers are held in tension when the diamond is deposited which is anticipated from the stress models. The estimated magnitude of the residual biaxial stress can be calculated using a linear frequency shift relation of 2.9 \cm{} GPa$^{-1}$\cite{Kuball2001}. Using the $E_2$(high) peak from the as-received GaN/III-N on Si wafer as the reference peak, an average tensile stress over various regions of the 0.5 and 2 mm membranes gave approximately 1.0 $\pm0.2$ GPa and 1.0 $\pm0.1$ GPa, respectively.  The magnified diamond peak region in Fig. \ref{fig:raman-zoom}.b shows minimal shift between the free-standing diamond film reference and at various regions on the grown sample. The free-standing film shows a marginal compressive shift to the value of 1332.8 $\pm0.02$ \cm{} from the typically observed value of 1332.5 \cm{}\cite{Ralchenko2011,Dychalska2015}. In contrast to the GaN/III-N spectra which all show tensile shifts from the reference GaN-on-Si wafer, the diamond peaks in both samples show varying global shifts from the reference film; compressive shifts for the 0.5 mm sample and tensile shifts for the 2 mm sample. These global shifts here show very small differences (around 0.2 GPa using a linear shift relation of 2.87 \cm{} GPa$^{-1}$\cite{Boppart1985}) in the intrinsic stresses in the deposited diamond.

\begin{figure}[t!]
    \setlength\tabcolsep{1pt}\centering
    \begin{tabular}[t]{cc}
	\includegraphics[width=0.34375\textwidth]{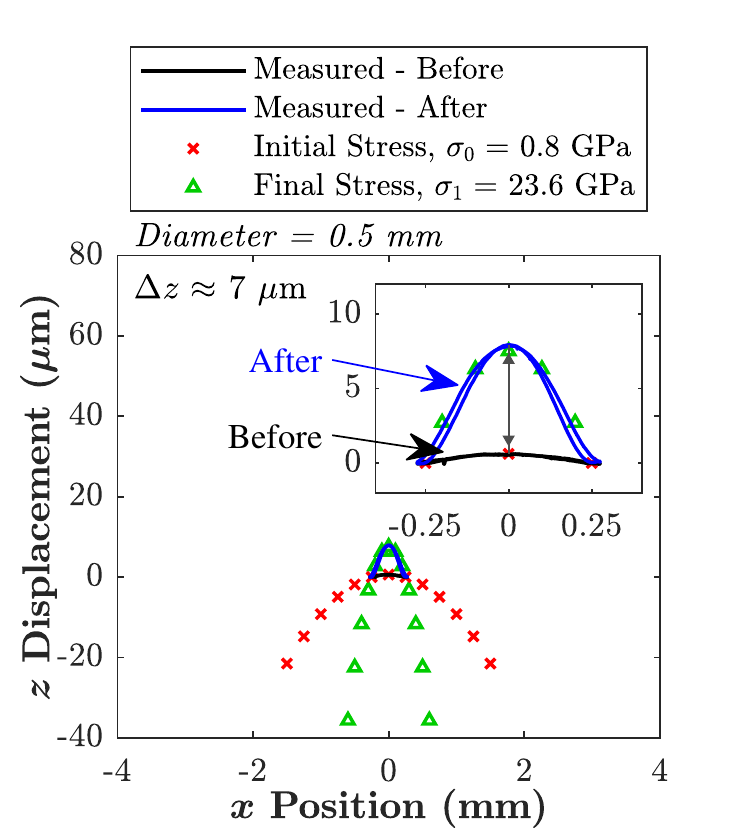} &
	\includegraphics[width=0.34375\textwidth]{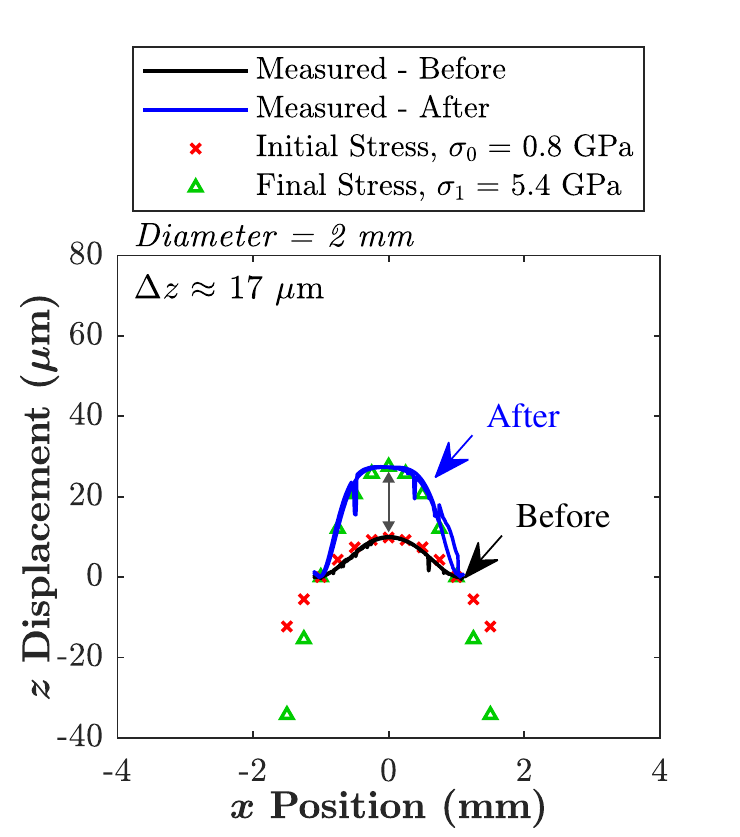} \\
	(a) & (b) \\
	\includegraphics[width=0.34375\textwidth]{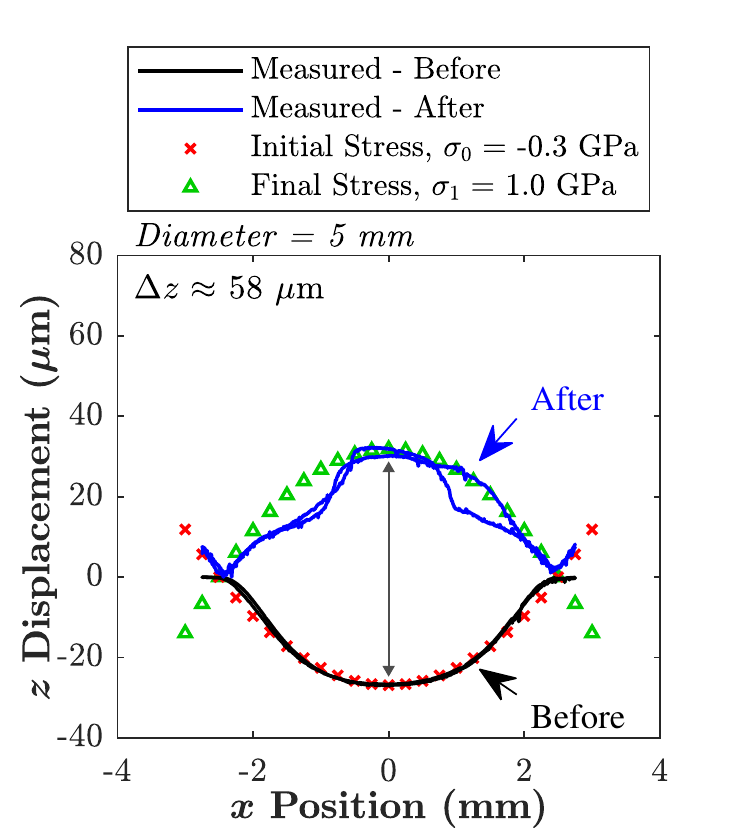} &
	\includegraphics[width=0.34375\textwidth]{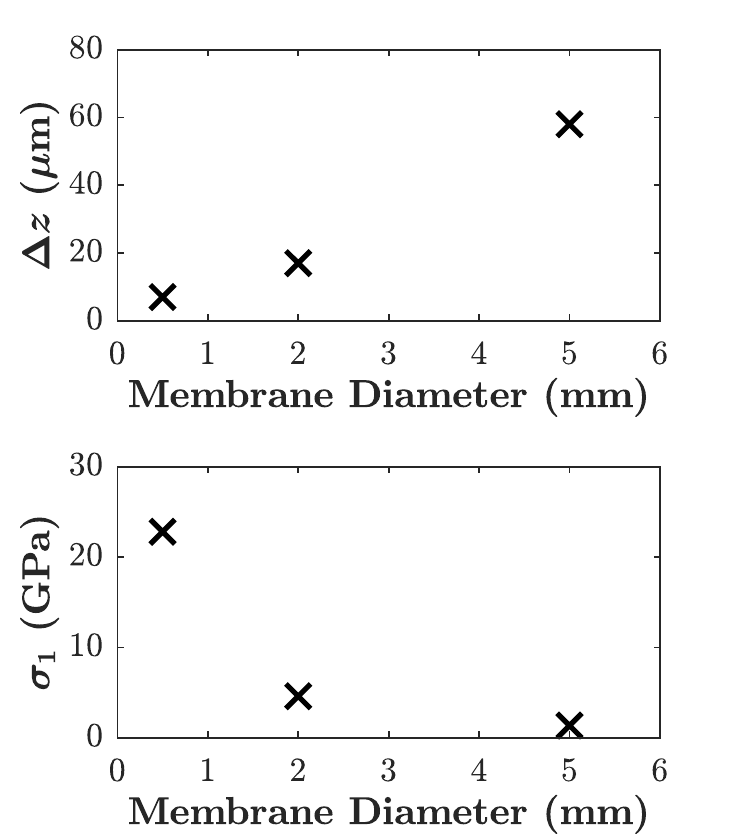} \\
	(c) & (d) 
    \end{tabular}
  \caption{Surface profilometry of GaN/III-N membranes before and after CVD diamond growth. (a) to (c) \CarbonC{show} two traces taken orthogonally \CarbonC{across the membrane}. \CarbonA{The bow ($\Delta z$) is the average of the displacement over the two orthogonal directions, with the largest standard deviation of \SI{2}{\micro\metre}}. The stress is estimated from Eq. (\ref{eq:stoney}) \CarbonC{using the curvature for the initial bow to obtain the initial stress} ($\sigma_0$) of a pre-stressed GaN film on Si \CarbonC{and the curvature for the final bow to obtain the final stress} ($\sigma_1$) for a GaN film on diamond. The stress is estimated using thicknesses: $t_{\textrm{GaN/III-N}}=$ \SI{5.3}{\micro\meter}, $t_{\textrm{Dia}}=$ \SI{50}{\micro\meter}, $t_{\textrm{Si}}=$ \SI{75}{\micro\meter}. (d) shows $\Delta z$ and $\sigma_1$ as a function of membrane diameter.}
  \label{fig:dektak}
\end{figure}

\subsection{Surface Profilometry}
The surface profilometry of the GaN side of the membranes before and after CVD growth is shown in Fig. \ref{fig:dektak}. In addition to the prediction of thermal stresses, the analytical model can also be used to infer the stresses using \RevBRS{Eq.} (\ref{eq:stoney}) and the defined radii of curvature. An indication as to whether a simplified analytical model can be used accurately or not can be determined by corroborating these values with the Raman spectroscopy measurements. A clear trend of increasing bow with diameter is found, however, large discrepancies are noted for the implied stress data. Starting with the smallest 0.5 mm diameter membrane, the initial bow was minimal at less than \SI{1}{\micro\metre}. After diamond growth, the bow was around \SI{7}{\micro\metre}. The calculated stress using \RevBRS{Eq.} (\ref{eq:stoney}) suggests an initial value of around 0.8 GPa and increases to a very large value of around \CarbonC{23.6} GPa after growth \CarbonC{, a value much higher than the experimentally measured tensile strength\cite{Brown2011}}. The 2 mm membrane showed a larger initial bow of \SI{10}{\micro\metre} and increased by \SI{17}{\micro\metre} after CVD diamond growth. The initial stress is similar at 0.8 GPa with a stress after growth of \CarbonC{5.4 GPa}, much lower than the 0.5 mm membrane. Finally, the 5 mm membrane shows a much larger initial strain in the opposite direction. While the initial strain is different, the stresses can be compared with the other samples based on the \textit{net} stress. Also, after diamond growth, a considerable difference is \RevDJW{found between} the orthogonal traces, with a large region warped  in the centre of the membrane. This is due to the fact that the large membrane has bowed significantly away from the heat sunk sample holder and pushed further into the plasma resulting in thermal runaway and damage to the GaN/III-N film. Using the initial curvature from the edges of the membrane, the maximum estimated net bow is considerably \RevDJW{larger} at around \SI{58}{\micro\metre}.

\RevCAR{
\section{Discussion}\label{sec:discuss}
There are several key findings of this thermal stress study that should be considered before designing and fabricating diamond on GaN/III-N membranes: (1) analytical and numerical modelling accuracy, (2) damage to larger membranes, (3) etching and redeposition of the Si border and finally (4) approaches to reducing membrane bow.

\subsection{Modelling accuracy}
The presented analytical model does not give an accurate reproduction of the experimental results; larger bow for large membranes, \CarbonC{overall smaller stresses than those found in experiment and most pertinent is the almost a non-existent bow for much smaller membranes}. This is indeed due to the additional stresses caused by the Si border that are not analytically modelled. The numerical model produces a better representation of the experimental data by exhibiting a \RevDJW{measurable} bow at smaller sizes, however, there still exists a discrepancy in larger sizes as with the analytical model. Fig. \ref{fig:fem-dia}, show bows of almost twice the experimental values, implying that the induced thermal stress is much higher in the model. 

One important factor that has not been considered in either of the models is the temperature gradient across the sample. In Fig. \ref{fig:photo}, it is clear that the corners of the sample are at much higher temperatures, owing to the bowing upwards of the edges into the plasma. The base thermal stress model assumes a uniform temperature across the whole sample. A simple linear spatial temperature gradient can be imposed to replicate this phenomenon that is scaled to the sample size through the following:
\begin{figure}[t!]
	\centering
	\includegraphics[width=0.45\textwidth]{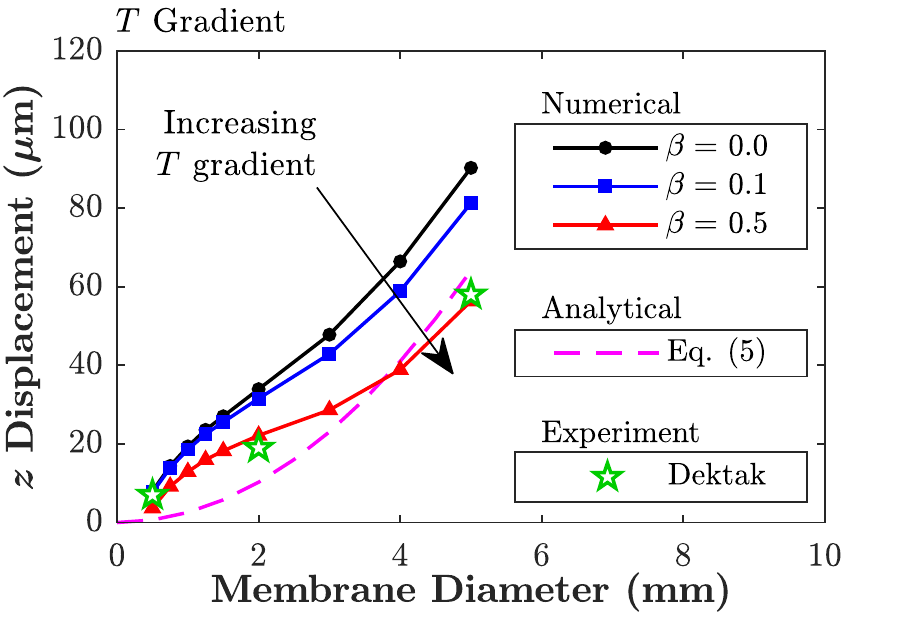}
	\includegraphics[width=0.45\textwidth]{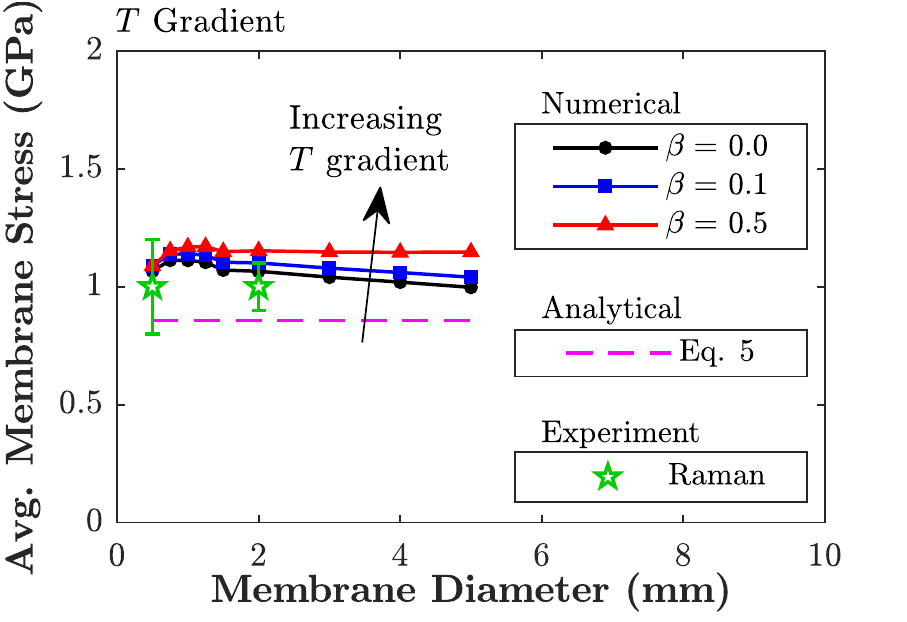}
	\caption{Comparison of membrane bow from experimental measurements, analytical model and numerical model with additional temperature gradients introduced. The numerical model parameters used were: $t_{\textrm{GaN}}=$ \SI{1.7}{\micro\meter}, $t_{\textrm{III-N}}=$ \SI{3.6}{\micro\meter}, $t_{\textrm{Dia}}=$ \SI{50}{\micro\meter}, $t_{\textrm{Si}}=$ \SI{75}{\micro\meter} and a deposition temperature of $T_g=720$ \degC{} and variation induced using Eq. \ref{eq:trad}. The analytical model uses the same constant deposition temperature and thicknesses with a combined GaN/III-N layer and the result showing the combined heating and cooling stresses.
	} 
	\label{fig:exp}
\end{figure}
 
\begin{equation}
	T(x,y) = T_g\left(1+\beta\frac{\sqrt{2(x^2+y^2)} }{W_{\textrm{sub}}} \right)
	\label{eq:trad}
\end{equation}

\noindent where $T$ is the spatially dependent temperature in \degC{}, $T_g$ is the growth temperature in the centre of the membrane, $\beta$ is a temperature variation parameter, $x$ and $y$ are the initial positions in Cartesian coordinates and $W_{\textrm{sub}}$ is the width of whole sample. Fig. \ref{fig:exp} demonstrates the use of a temperature gradient, which drastically reduces the bow in the centre to align with experimental values while increasing the stress by a small amount. For example, at a $\beta$ value of 0.5, the temperature linearly increases from 720 \degC{} at the centre to 1,080 \degC{} at the corner tip of the sample which in turn creates additional stresses which counteract the central bow. This demonstrates the importance of using numerical models for examining thermal stresses in GaN/III-N membrane structures.

\subsection{Heat sinking during MPCVD}
The damage to the largest membrane marks an interesting issue in the CVD diamond on GaN/III-N membrane approach. This comes down to a matter of bow and thermal runaway. Note that while the net bow of the 5 mm diameter membrane is \SI{58}{\micro\metre} its vertical displacement above the zero point  ($\sim$\SI{30}{\micro\metre})  is not that much larger than the 2 mm diameter membrane ($\sim$\SI{24}{\micro\metre}) which was not damaged. A plausible reason as to why the larger membrane sustained damage is that the heat sinking capability is less effective over larger areas owing to the poor thermal conductivity of GaN and AlGaN. The temperature at the membrane surface was likely much higher than for the other samples. This ultimately limits any large area CVD diamond growth over GaN/III-N membranes unless appropriate heat sinking is employed.

\subsection{Si etching and deposition}
The thin layer of SiO$_2$ at the interface between the CVD diamond and the N-polar side of the AlN is a potentially serendipitous finding. \CarbonA{This SiO$_2$ layer facilitates a stable carbide bond for the diamond to the III-N stack; a SiN bond with the N-polar side of the AlN and a SiC bond with the diamond. There are two cases where this layer may be deposited. The first is at the ICP membrane etch stage where the SiO$_2$ layer is simply due to an incomplete etch through the Si and the remaining few nanometres oxidise from exposure to air after the sample is brought up to atmosphere.  A second origin is during the MPCVD \nitro{}/\hydro{} plasma pre-treating stage before the seeding process where the Si border is etched and redeposited on the membrane. There are numerous studies on Si in \hydro{} microwave plasmas which demonstrate the production of the volatile silane (SiH$_4$) gas that diffuses to and reacts with a target substrate to redeposit Si\cite{Yamada2012a,Ohmi2009}. Pure SiH$_4$ as a source gas can also be used to co-deposit diamond onto substrates\cite{Sedov2019}. Once the sample is brought up to atmosphere after the \nitro{}/\hydro{} plasma pre-treatment, the deposited silicon oxidises to form a SiO$_2$ layer and the diamond seeds are introduced. It's also shown that the N-Polar face of AlN has been shown to have a high concentration of surface oxide groups\cite{Yoshikawa2016} which may contribute to the formation of SiO$_2$. From a stress perspective, the adhesion strength needs investigation as high stresses in this layer will result in delamination. This layer may be incredibly important for creating an interface with N-polar faces and CVD diamond, subject of another study\cite{Field2020}.}

\subsection{Reducing membrane bow}
Finally, the micron large membrane bows found for the CVD diamond on GaN/III-N membranes presents the biggest challenge for device manufacturing using contact lithography. Such large membrane bows cannot be dealt with as is, however, it is important to note that the results in this study demonstrate the \textit{net} membrane bows. This means that if the membrane is sufficiently pre-stressed before growth in the opposite direction, the bow upon heating will be counteracted and the CVD diamond would be deposited on a \textit{flat} GaN/III-N membrane. As per the model findings, the extreme stiffness of the diamond should then hold the membrane flat upon cooling, resulting in far less bow. This would also solve the heat sinking issues for larger membranes in point (2) as the membrane would not bow towards the plasma and become damaged. As a final processing step, the membranes can then be laser cut to reveal a free-standing GaN/III-N/diamond stack, ready for device processing. This idea of course requires significant GaN-on-Si wafer development to produce pre-stressed wafers, out of scope of this study.

\section{Conclusion}
The thermal stresses of the membrane approach for the integration of chemical vapour deposited (CVD) diamond with gallium nitride (GaN) on group-III-N (III-N) layers have been investigated using analytical and numerical modelling and experimental results. In utilising the Stoney analytical model in the presented way, the anticipated thermal stresses in GaN/III-N membrane structures are not accurately reproduced to the values found in experiment. While this is covered using numerical modelling, the result still requires careful consideration of the spatial temperature distribution over the sample whereby simple linear gradients appear to suffice. The dependence of various parameters including layer thicknesses, temperature and diamond stiffness on the membrane stress and bow has been modelled with the overall conclusion that the membrane deformation when it is heated defines the expected structure after cooling to room temperature. A CVD diamond layer as thin as \SI{1}{\micro\metre} is enough to lock this deformation in place, meaning that this initial deformation is the key to reducing bow. Experimental measurements using Raman spectroscopy corroborate the numerical models and additionally, simply using the radius of curvature from surface profilometry data to infer the stress is invalid at sizes lower than several millimetres. The use of a Si border in this membrane method also introduces SiO$_2$ at the interface between the diamond and the GaN/III-N stack. The membrane bow is the biggest issue that must be addressed as large membranes bow towards the microwave plasma and become damaged, however, if the membranes can be pre-stressed before growth, the final bow is expected to reduce}.

\section{Acknowledgements}
This project has been supported by Engineering and Physical Sciences Research Council (EPSRC) under program Grant GaN-DaME (EP/P00945X/1). J. A. Cuenca is an EPSRC Postdoctoral researcher. \RevDJW{DJ Wallis acknowledges support of EPSRC fellowship (EP/N01202X/2)}.

\bibliography{elsarticle_P2}

\end{document}